\documentclass[12pt]{revtex4-1}

\usepackage{graphicx}% Include figure files
\usepackage{dcolumn}% Align table columns on decimal point
\usepackage{amsmath}
\usepackage{color,soul}

\begin{document}

\title{Electrically Tunable Harmonics in Time-modulated Metasurfaces for Wavefront Engineering}% Force line breaks with \\

\author{Mohammad Mahdi Salary}
\affiliation{Metamaterial Laboratory, Electrical and Computer Engineering Department,
Northeastern University, Boston, Massachusetts 02115, USA
}%
\author{Samad Jafar-Zanjani}
\affiliation{Metamaterial Laboratory, Electrical and Computer Engineering Department,
Northeastern University, Boston, Massachusetts 02115, USA
}%

\author{Hossein Mosallaei}
\email{hosseinm@ece.neu.edu}
\affiliation{Metamaterial Laboratory, Electrical and Computer Engineering Department,
Northeastern University, Boston, Massachusetts 02115, USA
}%

\date{\today}% It is always \today, today,

%\begin{indented}
%\item[]August 2018
%\end{indented}

\begin{abstract}
Modulation of metasurfaces in time gives rise to several exotic space-time scattering phenomena by breaking the reciprocity constraint and generation of higher-order frequency harmonics. We introduce a new design paradigm for time-modulated metasurfaces, enabling tunable engineering of the generated frequency harmonics and their emerging wavefronts by electrically controlling the phase delay in modulation. It is demonstrated that the light acquires a dispersionless phase shift regardless of incident angle and polarization, upon undergoing frequency conversion in a time-modulated metasurface which is linearly proportional to the modulation phase delay and the order of generated frequency harmonic. The conversion efficiency to the frequency harmonics is independent of modulation phase delay and only depends on the modulation depth and resonant characteristics of the metasurface, with the highest efficiency occurring in the vicinity of resonance, and decreasing away from the resonant regime. The modulation-induced phase shift allows for creating tunable spatially varying phase discontinuties with 2$\pi$ span in the wavefronts of the generated frequency harmonics for a wide range of frequencies and incident angles. Specifically, we apply this approach to a time-modulated metasurface in the Teraherz regime consisted of graphene-wrapped silicon microwires. For this purpose, we use an accurate and efficient semi-analytical framework based on multipole scattering. We demonstrate the utility of the design rule for tunable beam steering and focusing of generated frequency harmonics giving rise to several intriguing effects such as spatial decomposition of harmonics, anomalous bending with full coverage of angles and dual-polarity lensing. Furthermore, we investigate the angular and spectral performance of the time-modulated metasurface in manipulation of generated frequency harmonics to verify its constant phase response versus incident wavelength and angle. The nonreciprocal response of the metasurface in wavefront engineering is also studied by establishing nonreciprocal links with large isolations via modulation-induced phase shift. The proposed design approach enables a new class of high-efficiency tunable metasurfaces with wide angular and frequency bandwidth, wavefront engineering capabilities, nonreciprocal response and multi-functionality.
\end{abstract}

\maketitle

\section{Introduction}

Optical metasurfaces have had a significant impact on development of novel optical devices owing to their capability in the engineering of electromagnetic wavefronts of light and their ultrathin structures which can reduce the footprint of optical platforms by replacing the bulky components. They have been used for numerous beam-shaping applications such as beam steering \cite{ni2012broadband,shi2014coherent,ni2013ultra}, focusing \cite{cheng2014wave, arbabi2015subwavelength} and holography \cite{walther2012spatial,ni2013metasurface}. These functionalities are realized by engineering the phase discontinuities across the metasurface according to the generalized Snell's law \cite{yu2011light, yu2014flat}. In conventional metasurfaces, generally two different approaches have been adopted for this purpose. In the first approach, the structural parameters of a resonant structure are varied across the metasurface which can provide a desired phase profile for linearly polarized light due to the phase agility at the resonance \cite{cheng2014wave, arbabi2015subwavelength}. In the second approach, a half-wave plate element is rotated around its axis which leads to a geometric phase shift equal to twice as the rotation angle for a circularly polarized incident wave based on the Parancharatnam-Berry (PB) design rule \cite{huang2012dispersionless,tymchenko2016advanced, forouzmand2016double}.

However, one major limitation associated with these conventional metasurfaces is their static optical response after fabrication which ties their application to the specific functionality they were designed for. To overcome this limitation, an extensive effort has been put into post-fabrication tuning of geometrically-fixed metasurfaces by exploiting mechanical \cite{zheludev2016reconfigurable, ou2013electromechanically}, thermal \cite{kim2016vanadium, tittl2015switchable}, optical \cite{abb2011all, li2014ultrafast} and electrical \cite{sherrott2017experimental,huang2016gate} effects which allows for realization of multiple functionalities and real-time engineering of light wavefronts. Among all the tunability mechanisms, electrical tunability of electro-optical materials via field-effect modulation is of particular interest as it yields a continuous tunability over a relatively wide range while possessing low power consumption and shorter response time compared to mechanical reconfiguration and thermal phase transition \cite{feigenbaum2010unity,lee2014nanoscale,cai2009compact}. Furthermore, unlike all-optical tunability, electrical tunability easily allows for  realization of graded-patterns through independent biasing of each element without requiring complex lens systems which makes it desirable for wavefront engineering. In particular, graphene and indium tin oxide (ITO) have attracted a lot of attention due to their compatibility with silicon technology, large scale fabrication feasibility and low-dimensionality of the active regions \cite{sherrott2017experimental,huang2016gate, park2016dynamic,kafaie2018dual}. The carrier concentration in these materials can be tuned through electrostatic gating in parallel capacitor configurations which can be translated into the change in the optical constants of the material through carrier-dependent dispersion models in the infrared (IR) and Terahertz (THz) frequencies. However, the change in the carrier concentration is limited to a thin accumulation layer for ITO and to a two-dimensional surface for graphene. Due to this fact, these materials have been incorporated in resonant geometries which can enhance the light-matter interaction in the active regions and enable tunable modulation of phase over a wide range in reflection  \cite{sherrott2017experimental, huang2016gate, forouzmand2016tunable, park2016dynamic, forouzmand2018tunable, kafaie2018dual} or transmission \cite{salary2017electrically} through spectral shift of the resonance. Despite the fruitful progress that has been made toward electrically tunable modulation of light wavefront, all the proposed designs thus far suffer from a limited efficiency as they cannot cover the entire 2$\pi$ span for phase modulation. Moreover, the afforded phase modulation comes at the cost of scarifying the amplitude efficiency which results into further decrement of efficiency. Furthermore, these metasurfaces exhibit strong resonant dispersion with extremely narrow bandwidth in both angular and spectral phase responses due to high quality-factor of the resonance required for tunability which makes them more susceptible to the fabrication errors and limits their functionality to a specific wavelength and incident angle. As such, realization of an electrically tunable metasurface covering the 2$\pi$ span for the phase modulation with wide spectral and angular bandwidths is a challenge yet-to-be addressed. In this contribution, we offer a new route for overcoming these limitations by introducing a dispersionless non-resonant phase shift of light upon undergoing temporal frequency transitions in time-modulated metasurfaces. 

Although electro-optical metasurfaces offer real-time tunability, their operation has been mostly studied in the quasi-static case where the temporal variations are ignored. The electro-optical materials such as graphene and ITO offer the opportunity to implement time-modulated metasurfaces by changing the external bias in time. Due to the quick response of these materials, they allow for modulation speeds up to several gigahertz \cite{phare2015graphene,wood2018gigahertz,babicheva2015transparent}. As such, they can be biased using radio-frequency (RF) circuits to realize time-modulated metasurfaces in IR and THz frequencies. Introducing time-modulation in a structure leads to generation of higher-order frequency harmonics and breaks the time-reversal symmetry \cite{zurita2009reflection,zurita2010resonances,shaltout2015time,liu2016time}. A significant effort has been devoted into developing magnetless nonreciprocal devices by adopting a periodic spatiotemporal modulation along the propagation direction which mimics a traveling wave motion and enables isolation and circulation of light by imparting different momentums to forward and backward propagating waves \cite{sounas2017non,lira2012electrically,hadad2015space,hadad2016breaking,correas2016nonreciprocal,taravati2017nonreciprocal, taravati2017mixer, chamanara2017optical,taravati2017nonreciprocal2,shi2017optical}. Moreover, time-modulated metasurfaces hold a great potential to extend the degree of light manipulation. It has been demonstrated that space-time photonic transitions of guided modes into leaky-modes in spatiotemporally-modulated metasurfaces can enable dynamic nonreciprocal beam-scanning \cite{hadad2015space,taravati2017mixer,shi2017optical}. Such a paradigm has a narrow operational bandwidth and is ill-suited for more advanced functionalities. In this work, we introduce a new paradigm for wavefront engineering using time-modulated metausrfaces which is based on modulation-induced phase shift of light when the constituent elements are modulated with a temporal phase delay. A very recent work has illustrated this concept experimentally in the RF regime using a time-modulated Huygens metasurface consisted of electric and magnetic dipole unit cells  \cite{liu2018huygens}. In particular, the independent control over the modulation of electric and magnetic building blocks has been leveraged to also control the directionality and frequency conversion efficiency of the metasurface \cite{liu2018huygens}.

Here, we establish modulation-induced phase shift of light as the basis of a general design paradigm for time-modulated metasurfaces while highlighting its great promise for realization of broadband and wide-angle metasurfaces capable of manipulating the higher-order frequency harmonics and shaping their wavefronts with electrical tunability. We demonstrate that by changing the phase delay in the temporal modulation of a metasurface using phase shifters, each generated frequency harmonic will acquire a dispersionless phase shift linearly proportional to its corresponding order covering 2$\pi$ span while maintaining a constant amplitude, independent of incident angle and polarization. The frequency conversion efficiency of the metasurface is determined by the geometry of constituent building blocks and its resonant characteristics, exhibiting a peak in proximity of the resonance, and decreasing by moving away from the resonant regime. The design rule is established using an analytical model for a general homogenized metasurface consisted of polarizable dipolar particles. The applicability of the design rule is subsequently studied in-depth and is verified beyond the dipolar regime by considering a time-modulated metasurface in THz regime based on graphene-wrapped microwires. For this purpose we use a rigorous multipole scattering technique recently developed by authors which allows for accurate and efficient characterization of time-modulated metasurfaces with a large difference between time scales of optical and modulation frequencies \cite{salary2018time}. The steering and focusing of higher-order frequency harmonics are demonstrated and the associated exotic scattering phenomena such as spatial decomposition of temporal frequency harmonics, anomalous bending and dual-polarity lensing are discussed. Moreover, the spectral and angular performance of functional time-modulated metasurfaces are investigated to verify the invariance of modulation-induced phase shift by changing the wavelength and incident angle. The modulated-induced phase shift is also leveraged to establish nonreciprocal links with large isolation.

\section{Design Rule: Modulation-induced Phase Shift of Light}
In order to simplify the mathematical analysis and gain insight to the scattering of light from time-varying metasurfaces, we utilize the equivalent surface admittance representation in this section to establish the design rule for a general time-modulated metasurface through analytical relations for reflection and transmission coefficients. In the subsequent sections, we verify the design rule for a time-modulated metasurface with realistic structure implemented by graphene-wrapped microwires as building blocks, analyzed without any approximations using a robust semi-analytical framework based on multipole scattering theory developed in \cite{salary2018time}. 

Let us consider a periodic array consisted of subwavelength elements varying in time. We assume the size of elements is small enough comparing to the wavelength, so that the array can be modeled with an effective time-varying surface admittance \cite{hadad2015space}. The effective admittance is directly related to the dipolar polarizability of the inclusions forming the array which is determined by their shape and constituent materials and can be retrieved numerically through calculation of scattering parameters \cite{zhao2011homogenization}. We emphasize that this homogenization model is valid only for deeply subwavelength elements in a periodic arrangement and it has not been used for the analysis in the following sections.

\begin{figure}[htbp]
\centering
{\includegraphics[width=2.5in]{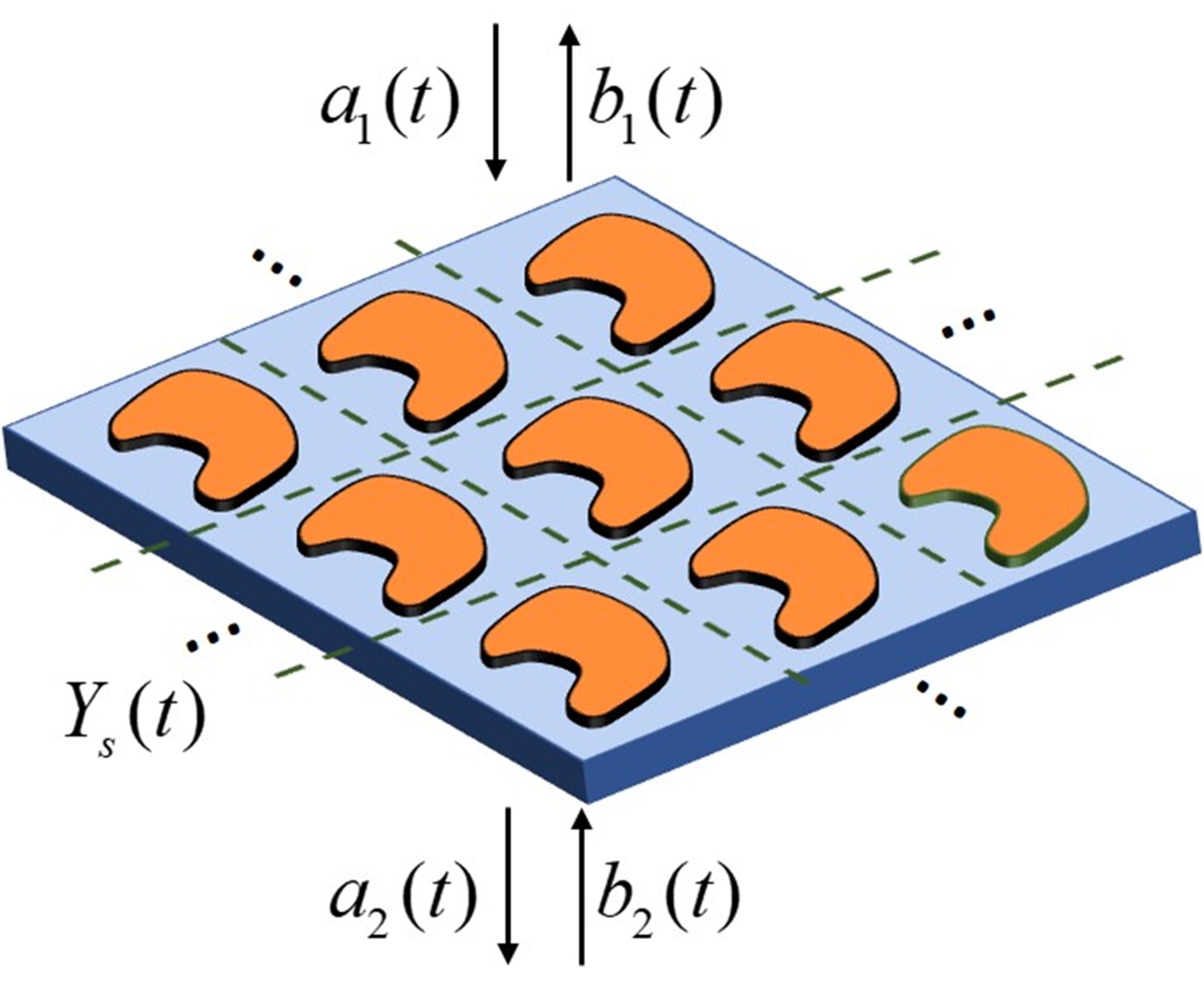}}
\caption{
The schematic of a general time-varying metasurface consisted of a 2D array of subwavelength elements modulated in time, modeled by an equivalent surface admittance.}
\label{fig:Fig1}
\end{figure}

Extending the well-known transfer matrix formalism, the tangential electromagnetic fields in both sides of the surface admittance are expanded into backward and forward propagating plane waves with time-varying complex amplitudes $a_m(t)$ and $b_m(t)$ as shown in Fig. \ref{fig:Fig1}.
 Considering the achievable modulation frequencies of electro-optical materials are limited to several GHz which are significantly smaller than the optical frequencies in THz and IR frequency regimes, the modulation-induced dispersion effects are indeed negligible. In such a case, the transfer matrix equation relating the amplitudes of the fields on opposite sides of the surface admittance can be expressed in multiplicative form in time-domain for each excitation frequency ($\omega_0$) as \cite{zhan2013transfer,mousavi2014strong}:
\begin{equation}
\label{eq:refname1}
\begin{bmatrix}
a_1(t)\\
b_1(t)
\end{bmatrix} =\overline{\overline{T}}(t)\begin{bmatrix}
a_2(t)\\
b_2(t)
\end{bmatrix} =
\begin{bmatrix}
1+Y_s(\omega_0,t)/2Y_0 &  Y_s(\omega_0,t)/2Y_0\\
-Y_s(\omega_0,t)/2Y_0 & 1-Y_s(\omega_0,t)/2Y_0 
\end{bmatrix}
\begin{bmatrix}
a_2(t)\\
b_2(t)
\end{bmatrix}
\end{equation}
where $\overline{\overline{T}}(t)$ is the time-varying transfer matrix, $Y_s(\omega_0,t)$ is the time-varying surface admittance of the metasurface and $Y_0$ is the admittance of plane wave in free-space which depends on the incident angle and polarization. Here, we do not set any limit on $Y_0$, so that the results are valid regardless of incident angle and polarization. It should be emphasized that while neglecting the modulation-induced dispersion effects, we include dependency of admittance temporal profile to the excitation frequency ($\omega_0$), thus accounting for the change in material parameters with optical frequency.

The transfer matrix equation can be taken into angular frequency domain by taking the Fourier transform of both sides as:
\begin{equation}
\begin{bmatrix}
A_1(\omega)\\
B_1(\omega)
\end{bmatrix}=
\overline{\overline{T}}(\omega_0,\omega)*
\begin{bmatrix}
A_2(\omega)\\
b_2(\omega)
\end{bmatrix}=\int \overline{\overline{T}}(\omega_0,\omega-\omega')\begin{bmatrix}
A_2(\omega')\\
b_2(\omega')
\end{bmatrix}d\omega'
\label{eq:refname1}
\end{equation}

where $A_{1,2}(\omega)$ and $B_{1,2}(\omega)$ are the complex amplitudes of forward and backward propagating plane waves at the opposite sides of the metasurfaces in the frequency domain, respectively, and * denotes convolution. Equation (2) implies that an input frequency $\omega_0$ will be converted to a spectrum of output frequencies. When the elements are varying with a periodic modulation having a modulation frequency of $\omega_m$, the transfer matrix is also periodic and can be expanded in form a Fourier series as:
\begin{equation}
\overline{\overline{T}}(t)=\sum_{n}\overline{\overline{T}}^n(\omega_0)\exp(in\omega_mt)
\end{equation}
And the Fourier transform takes the following form:
\begin{equation}
\overline{\overline{T}}(\omega)=\sum_{n}\overline{\overline{T}}^n(\omega_0)\delta(\omega-n\omega_m)
\end{equation}
Therefore, the generated frequency harmonics will be a discrete spectrum consisted of the input frequency up- and down-modulated by the time modulation frequency. 

Substituting Eq. (4) into (2) while choosing $\omega=\omega_q=\omega_0+q\omega_m$ and $\omega'=\omega_p=\omega_0+p\omega_m$ with $q,p=\cdots,-1,0,+1,\cdots$, we will arrive at the following system of equations:
\begin{equation}
\Bigg\{ \begin{bmatrix}
A_1(\omega_q)\\
B_1(\omega_q)
\end{bmatrix}\Bigg\}=
\big\{\overline{\overline{T}}^{q-p}(\omega_0)\big\} \Bigg\{ \begin{bmatrix}
A_2(\omega_p)\\
B_2(\omega_p)
\end{bmatrix}\Bigg\}
\label{eq:refname1}
\end{equation}

The above matrix equation relates the complex amplitudes of the forward and backward propagating frequency harmonics on opposite sides of the metasurface in the frequency domain. By setting $B_2(\omega_p)=0$ for all $p$'s and $A_1(\omega_q)=0$ for all $q$'s except $A_1(\omega_0)=1$, we can solve the equation to obtain the reflection and transmission coefficients of all generated frequency harmonics for a monochromatic excitation of $\omega_0$ incident from the top, as:
\begin{align}
r(\omega_0+n\omega_m)&=\frac{B_1(\omega_0+n\omega_m)}{A_1(\omega_0)}\\
t(\omega_0+n\omega_m)&=\frac{A_2(\omega_0+n\omega_m)}{A_1(\omega_0)}
\end{align}

By assuming the time-modulation profile as a simple offset sinusoid, the surface admittance of the metasurface can be expressed as:
\begin{equation}
\begin{split}
Y_s&(\omega_0,t)=Y_s^{0}(\omega_0)(1+\Delta Y_s\cos(\omega_mt))\\
&=
Y_s^{-1}(\omega_0)\exp(-i\omega_mt)+Y_s^{0}(\omega_0)+Y_s^{+1}(\omega_0)\exp(+i\omega_mt)
\end{split}
\end{equation}
where $Y_s^{-1}(\omega_0)=Y_s^{+1}(\omega_0)=\frac{Y_s^{0}(\omega_0)\Delta Y_s}{2}$. In the above relation $Y_s^{0}(\omega_0)$ is the average surface admittance and $\Delta Y_s$ is the admittance modulation depth which depends on the resonant characteristics of the building block. Substituting the admittance into the time-varying transfer matrix and comparing with Eq. (3), we will obtain the Fourier coefficients of the transfer matrix as:
\begin{align} 
& \overline{\overline{T}}^{-1}(\omega_0)=\frac{Y_s^{-1}(\omega_0)}{2Y_0}\begin{bmatrix}
1 &  1\\
-1 & -1 
\end{bmatrix}\\
& \overline{\overline{T}}^{0}(\omega_0)=\begin{bmatrix}
1+Y_s^0(\omega_0)/2Y_0 &  Y_s^0(\omega_0)/2Y_0\\
-Y_s^0(\omega_0)/2Y_0 & 1-Y_s^0(\omega_0)/2Y_0 
\end{bmatrix}
\\
& \overline{\overline{T}}^{+1}(\omega_0)=\frac{Y_s^{+1}(\omega_0)}{2Y_0}\begin{bmatrix}
1 &  1\\
-1 & -1 
\end{bmatrix}
\end{align}
Plugging (9)-(11) into (5) and truncating the frequency harmonics into $\pm M_f$ results into a tri-block diagonal matrix equation consisted of $2(2M_f+1)$ equations which can be readily solved for $A_2$'s and $B_1$'s by setting $\forall	p,\forall q\neq0 \Rightarrow  B_2(\omega_p)=A_1(\omega_q)=0$ and $A_1(\omega_0)=1$ to obtain the transmission and reflection coefficients of the generated frequency harmonics as given in (6) and (7).

Now, let us consider adding a phase delay of $\alpha$ to the sinusoidal modulation profile as $Y_s(\omega_0,t)=Y_s^0(\omega_0)(1+\Delta Y_s \cos(\omega_mt+\alpha))$. The Fourier coefficients of the transfer matrix will be subsequently changed as $\overline{\overline{T}}_\alpha^{\pm 1}(\omega_0)=\exp(\pm i\alpha) \overline{\overline{T}}^{\pm 1}(\omega_0)$ and $\overline{\overline{T}}_\alpha^{0}(\omega_0)= \overline{\overline{T}}^{0}(\omega_0)$, where subscript $\alpha$ denotes the quantities corresponding to the time-modulation with phase delay of $\alpha$. Writing transfer matrix equation (5) for the phase-delayed modulation and using $\overline{\overline{T}}_\alpha^{q-p}(\omega_0)= \exp( i(q-p)\alpha)\overline{\overline{T}}^{q-p}(\omega_0)$, we have:
\begin{equation}
\Bigg\{ \begin{bmatrix}
A_{1,\alpha}(\omega_q)\\
B_{1,\alpha}(\omega_q)
\end{bmatrix}\Bigg\}=
\big\{\exp( i(q-p)\alpha)\overline{\overline{T}}^{q-p}(\omega_0)\big\} \Bigg\{ \begin{bmatrix}
A_{2,\alpha}(\omega_p)\\
0
\end{bmatrix}\Bigg\}
\label{eq:refname1}
\end{equation}
In the above equation, we have set $B_{2,\alpha}(\omega_p)=0$ for all $p$'s and $A_{1,\alpha}(\omega_q)=0$ for all $q$'s except $A_{1,\alpha}(\omega_0)=1$ to obtain the transmission and reflection coefficients corresponding to the phase-delayed modulation. Defining $c_p=\frac{A_{2,\alpha}(\omega_p)}{A_2(\omega_p)}$, one can easily obtain:
\begin{align}
\begin{bmatrix}
A_{2,\alpha}(\omega_p)\\
0
\end{bmatrix}&=c_{p}\begin{bmatrix}
A_2(\omega_p)\\
0
\end{bmatrix}\\
\begin{bmatrix}
A_{1,\alpha}(\omega_q)\\
B_{1,\alpha}(\omega_q)
\end{bmatrix}&=c_{p}\exp(i(q-p)\alpha)\begin{bmatrix}
A_1(\omega_q)\\
B_1(\omega_q)
\end{bmatrix}
\end{align}
which results into:
\begin{align}
r_\alpha(\omega_0+n\omega_m)&=\frac{B_{1,\alpha}(\omega_0+n\omega_m)}{A_{1,\alpha}(\omega_0)}=\frac{c_{0}\exp(in\alpha)B_1(\omega_n)}{c_{0}A_1(\omega_0)}=\exp(in\alpha)r(\omega_0+n\omega_m)\\
t_\alpha(\omega_0+n\omega_m)&=\frac{A_{2,\alpha}(\omega_0+n\omega_m)}{A_{1,\alpha}(\omega_0)}=\frac{c_{n}A_2(\omega_n)}{c_{n}\exp(-in\alpha)A_1(\omega_0)}=\exp(in\alpha)t(\omega_0+n\omega_m)
\end{align}
These equations imply that by changing the modulation phase delay ($\alpha$), the generated frequency harmonic of order $n$ will acquire a phase shift of $n\alpha$ in forward and backward scattering planes while maintaining a constant amplitude. This provides a span of $2n\pi$ for the phase shift of $n$-th frequency harmonic by changing $\alpha$ from 0 to $2\pi$ which can be used as a design principle for wavefront engineering of generated frequency harmonics. Furthermore, according to the formulation, the modulation-induced phase shift occurring upon frequency conversions in time-modulated metasurfaces is non-resonant, dispersionless and independent of incident angle and polarization. Due to the conjugate phase of up-and down-modulated harmonics and the difference between the acquired phase of different harmonic orders, this design principle leads to several exotic phenomena such as anomalous bending, spatial decomposition of frequency harmonics and dual-polarity focusing/diffusing. In this sense, the proposed design rule has a close resemblance to the PB design rule in which a circularly-polarized light undergoing polarization conversion in a half-wave plate acquires a dispersionless geometric phase shift by rotating elements where the phase shift is conjugate for circular polarizations of opposite handness \cite{huang2012dispersionless,tymchenko2016advanced}.

The established design rule allows for realization of advanced functionalities through electrically phased time-modulated metasurfaces implemented by using electro-optical materials such as graphene and ITO. The optical constants of these materials can be modulated with modulation frequencies up to several GHz which can be achieved by using a radio-frequency (RF) biasing network. The phase delay in modulation can be realized and electrically tuned using RF phase shifters. It should be remarked that the phase shifting of electrical modulation is sufficient instead of a true time-delay. Multiple techniques can be adopted to realize passive and active RF phase shifter circuits \cite{firouzjaei2010mm} which have been extensively used in phased array microwave antennas. It should be noted that conventional phased-array antennas are time-invariant and utilize RF phase shifters in their feeding network to generate a phased radiation of fundamental frequency harmonic. Whereas, a phased time-modulated metasurface relies on an optical excitation (plane wave incidence) for radiation and the scatterers are modulated in time with an RF electrical network to engineer the emerging wavefronts of higher-order frequency harmonics.

The developments made in this section are agnostic to the shape and material composition of the building blocks; implying that the design rule can be implemented with arbitrary-shaped unit cells incorporating different types of tunable materials in different frequency regimes. In the following, we will synthesize a time-modulated metasurface in the THz regime using graphene-wrapped silicon microwires and apply this design rule to verify its validity beyond the dipolar regime while exploring beam-shaping and exotic phenomena in more depth. All the results in the following sections are obtained with an exact semi-analytical method based on multipole scattering developed in \cite{salary2018time} without using equivalent admittance approximation.

\section{Time Modulated Metasurface}
\subsection{Implementation based on Graphene-wrapped Microwires}
Graphene has been extensively used as a tunable 2D plasmonic material in the mid-IR and THz frequencies. The surface conductivity of graphene can be tuned over a wide range by electrical biasing in a parallel capacitor configuration which leads to a change in the Fermi energy level via field-effect. A monolayer graphene sheet can support highly confined surface plasmon polaritons (SPPs) which provides an ideal platform for tunable manipulation of guided light at nanoscale \cite{woessner2015highly,lin2017all,jiang2018group}. Moreover, incorporation of graphene in resonant structures can yield a large tunability in the optical response of metasurfaces via enhanced light-graphene interactions. It has been shown that graphene-integrated metal-insulator-metal configurations can be used to modulate the absorption \cite{yao2014electrically} and to tune the reflection phase over a wide range \cite{yao2014electrically} at the cost of low amplitude and narrow bandwidth \cite{sherrott2017experimental}. In the account of its flexibility, graphene has also been used in conformal geometries such as graphene-wrapped wires which can enhance light-matter interactions via Mie resonant modes. The fabrication feasibility of such structures via chemical-vapor-deposition growth of monolayer graphene \cite{wang2010large, wu2014graphene} or draping graphene flakes over wires with an adhesive tape \cite{li2014ultrafast} has been demonstrated. Furthermore, they have been exploited for various applications such as tuning the absorption \cite{zhang2016electrically}, cloaking \cite{chen2011atomically, forouzmand2015electromagnetic}, waveguiding \cite{gao2014single, zhang2015tunable, davoyan2016salient} and nonlinear harmonic generation \cite{gao2016second}.

The modulation of Fermi energy level of graphene has been experimentally demonstrated with modulation frequencies up to 30 GHz \cite{phare2015graphene}. As such, it is an excellent candidate for realization of time-modulated metasurfaces. In Ref. \cite{correas2016nonreciprocal}, spatiotemporal modulation of graphene sheets has been proposed to develop power isolator waveguides and nonreciprocal leaky-wave antennas through space-time modal transitions of SPPs. In Ref. \cite{salary2018time}, the authors have demonstrated theoretical realization of space-time gradient metamaterials based on graphene-wrapped microwires by using a robust semi-analytical framework based on multipole scattering which overcomes the challenges facing time-domain simulation techniques for accurate characterization of time-modulated systems with significantly different time-scales of optical and modulation frequencies. Moreover, the potential applications of time-modulated metasurfaces and metamaterials in holographic generation of frequency harmonics and spatiotemporal manipulation of light are briefly explored. In this work, we apply the developed modeling framework in \cite{salary2018time} to a similar time-modulated metasurface architecture for an in-depth analysis of modulation-induced phase shift in time-modulated metasurfaces and establishing its dependence on the harmonic order and its independence from the incident wavelength, angle and polarization. It should be remarked that the choice of this specific architecture have been due to the robustness and efficiency of the semi-analytical multipole scattering technique. This allows us to bring out the essential physics and illustrate the rich behavior of time-modulated metasurfaces through comprehensive studies which would have been prohibitively challenging using other computational methods. The concepts and underlying physics can be extended to simpler active optical elements more suited for fabrication. 

\begin{figure}[htbp]
\centering
{\includegraphics[width=12cm]{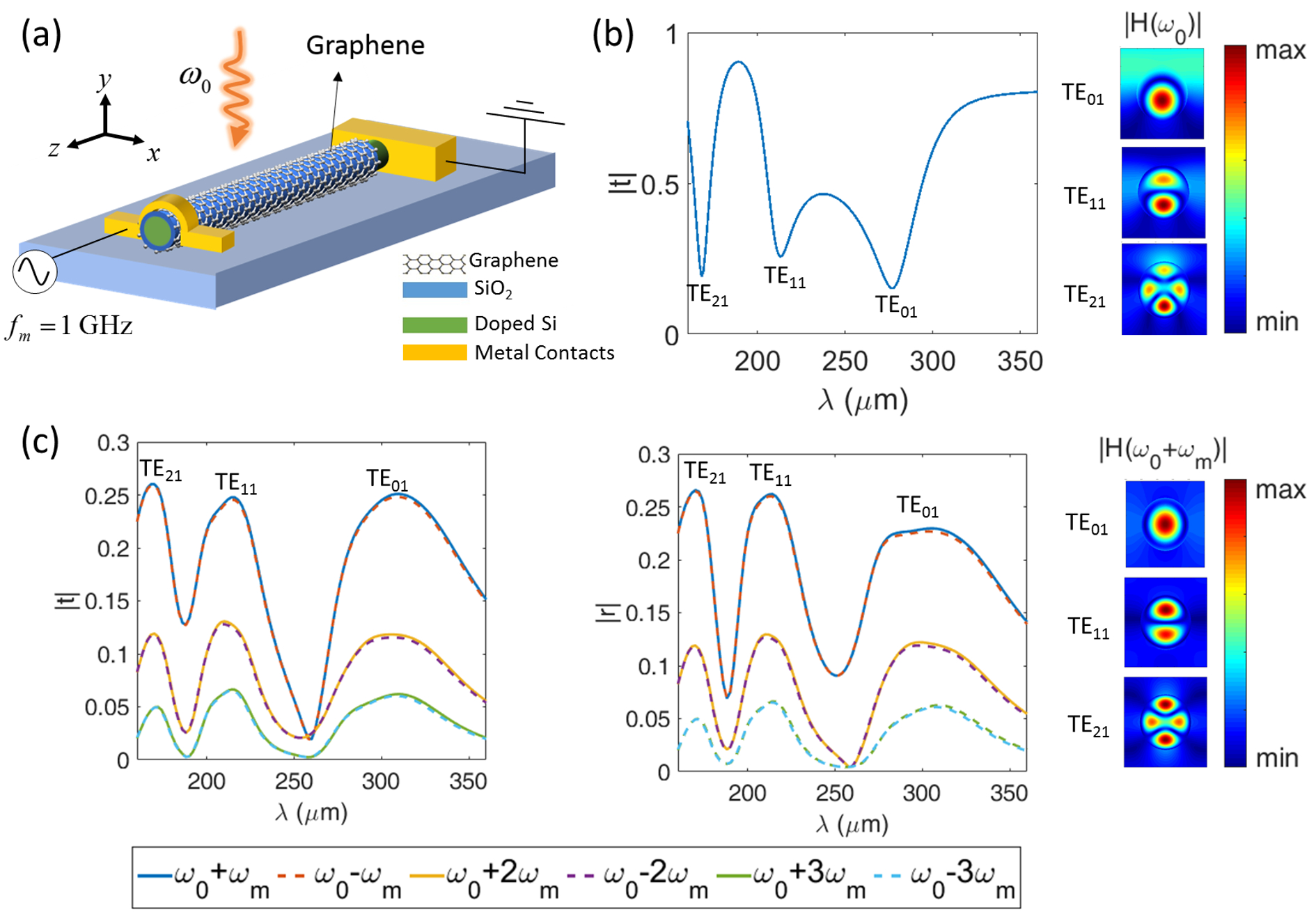}}
\caption{(a) The unit cell of implemented time-modulated metasurface by adopting graphene-wrapped Si (30$\mu$m)/SiO$_2$ (20 nm) microwires as building blocks with a periodicity of $\Lambda=130$ $\mu$m. (b) The amplitude of transmission coefficient of the metasurface obtained as a function of wavelength for the un-modulated case with $E_{f0}=0.3$ eV for a normally incident TE polarized plane wave. The dips in the transmission coefficient amplitude correspond to the resonant modes which are characterized according to the magnetic field profiles. (c) The calculated amplitude of transmission and reflection coefficients corresponding to the generated frequency harmonics by the time-modulated metasurface with $E_{f0}=0.3$ and $\delta=0.85$ as functions of incident wavelength. The nearfield distributions demonstrate the magnetic field profiles of the first-order generated frequency harmonic at the resonant wavelengths.}
\label{fig:Fig2}
\end{figure}

The unit cell of the implemented metasurface is demonstrated schematically in Fig. \ref{fig:Fig2}(a). It is composed of an array of subwavelength microwires with a periodicity of $\Lambda=130$ $\mu$m on a silica substrate with a thickness of $80$ $\mu$m. The topology of microwires are chosen such that they support Mie resonances to increase modulation depth and frequency conversion efficiency, and also allow for biasing graphene in a parallel capacitor configuration. The cores are made of silicon with radius of $R_{core}=40$ $\mu$m which are subsequently coated by thin insulating SiO$_2$ layers with radial thickness of 20 nm, in accordance to the biasing requirements. The geometrical parameters of the metasurface unit-cell are chosen for operation in the low-THz regime aiming for high frequency conversion in the operating band through engineering the resonant characteristics while minimizing the undesirable diffractions and coupling between the neighbor elements. Further discussion on the geometrical parameters and parametric studies are brought in section 1 of Supplemental Material. The electrical biasing of the element can be done by applying a voltage between the silicon core and the graphene which requires doping of the silicon \cite{sorger2012ultra}. Here, we use a moderately doped n-type silicon with the background carrier concentration of $n=10^{15}$ cm$^{-3}$. The silicon cores are connected to a ground plane while a time-varying electrical voltage is applied to the graphene layers through a 2D biasing grid which allows addressing and biasing each element, independently. The proposed configuration is inspired by field-effect transistors using graphene-wrapped silicon microwires \cite{jin2011graphene}.

The complex permittivity of silicon is expressed in terms of a Drude model \cite{naik2013alternative} as $\epsilon(\omega)=\epsilon_{inf}-\frac{ne^2}{\epsilon_0m^*}\frac{1}{\omega^2+i\omega\Gamma}$ in which $\epsilon_{inf}=11.7$ is the high-frequency permittivity, $\Gamma=81.426\times10^{12}$ rad/s is the damping constant, $n$ is the carrier concentration, $e$ is the electron charge, $\epsilon_0$ is the vacuum permittivity and $m^{*}=0.27m_e$is the effective mass of electron charge carrier. The dispersive surface conductivity of graphene in the THz regime is dominated by intraband electronic transitions which can be approximated as following, for a moderately doped graphene near the room temperature \cite{rakheja2016gate}:
\begin{equation}
\sigma_{intra}(\omega)\approx\frac{2e^2}{\pi \hbar^2} \frac{k_BT}{\omega+i\tau^{-1}}(\frac{E_f}{2k_BT})
\end{equation}
where $e$ is the electron charge, $\hbar$ is the reduced Planck's constant, $k_B$ is the Boltzmann constant, $T$ is temperature ($k_BT=25.7$ meV at room temperature), $\tau$ and $E_f$ are the scattering time and Fermi energy level of graphene, respectively. The scattering time is related to the Fermi energy level, Fermi velocity and carrier mobility. Here, we have fixed the relaxation time as $\tau=0.5$ ps corresponding to a high quality monolayer graphene \cite{woessner2015highly} and disregarded the interplay of Fermi energy level and scattering time.

The Fermi energy level can be controlled by biasing graphene in a parallel capacitor configuration and applying an exterior gate voltage. The dependence of the Fermi energy to the voltage is determined by the capacitance of gate dielectric as well as quantum capacitance of graphene \cite{rakheja2016gate,jin2011graphene}. Studying such effects has been out of the scope of this work and we assume a sinusoidal modulation profile is achieved for Fermi energy level without making any assumptions on the required waveform of the biasing signal. According to Eq. (17), a temporal modulation of Fermi energy level linearly translates to a similar modulation profile for graphene's conductivity. Applying a gate voltage with a sine squared waveform such that $E_f=E_{f0}(1+\delta \sin(\omega_mt+\alpha))$, the temporal variation of graphene conductivity for an excitation frequency of $\omega_0$ can be written as:
\begin{equation}
\sigma(\omega_0,t)=\sigma_{intra}(\omega_0,E_{f0})(1+\delta\sin(\omega_mt+\alpha))
\end{equation}
Recent experimental demonstrations have reported that the Fermi energy level of graphene can be modulated in the range of 0-0.6 eV using electrical gating \cite{sherrott2017experimental}. Since the frequency conversion efficiency in time-modulated metasurfaces is proportional to the modulation depth \cite{salary2018time}, here we consider $E_{f0}=0.3$ eV and $\delta=0.85$ to achieve largest modulation depth while remaining in the moderately doped regime of graphene where surface conductivity can be expressed as Eq. (18).

The calculated transmission coefficient spectrum of the metasurface in the un-modulated case with a Fermi energy level of $E_{f0}=0.3$ eV is shown in Fig. \ref{fig:Fig2}(b) for a normally incident plane wave with transverse electric (TE) polarization in which the magnetic field is along microwire axis. The dips in the transmission coefficient correspond to the resonant modes which can be characterized according to their magnetic field mode profiles shown in the figure (in TE$_{mn}$, $m$ denotes the azimuthal and $n$ denotes the radial mode index). TE$_{01}$ and TE$_{11}$ and TE$_{21}$ modes can also be referred to as magnetic dipole (MD) and electric dipole (ED) and electric quadropole (EQ) modes, respectively \cite{petschulat2010understanding}. The sinusoidal modulation of graphene conductivity in time will lead to generation of higher-order frequency harmonics $\omega_0\pm n\omega_m$ for a monochromatic excitation frequency of $\omega_0$. The modulation frequency is chosen as $f_m=1$ GHz which is well within the practical range \cite{yao2014electrically, liu2011graphene} while also allows for resolving the generated frequency harmonics with the resolution of current THz detectors and spectrometers \cite{yasui2006terahertz}. The transmission and reflection coefficients of the frequency harmonics generated by the time-modulated metasurface in the non-diffractive regime are defined as $t(\omega_0+n\omega_m)=\frac{H_{f}(\omega_0+n\omega_m)}{H_i(\omega_0)}$ and $r(\omega_0+n\omega_m)=\frac{H_{b}(\omega_0+n\omega_m)}{H_i(\omega_0)}$ in which $H_i$ is the incident magnetic field while $H_{s,f}$ and $H_{s,b}$ are the magnetic fields corresponding to the plane waves propagating in forward and backward directions, respectively. The amplitudes of transmission and reflection coefficients corresponding to the higher-order frequency harmonics are obtained and plotted in Figs. \ref{fig:Fig2}(c) and (d) as functions of incident wavelength, respectively. As it can be noted from the results, the higher-order frequency harmonics are generated in both forward and backward directions due to geometrical symmetry of the modulated scatterer and non-overlapping magnetic and electric responses \cite{liu2018huygens}. The highest frequency conversion efficiencies are obtained in the vicinity of resonant wavelengths of fundamental frequency harmonic due to the larger scattering modulation depth and stronger light-graphene interaction \cite{salary2018time}. It should be noted that in the account of modulation frequency being small with respect to the optical frequency, the amplitudes of up- and down-modulated harmonics are almost equal. The modulation-induced phase shift is independent from the modulation frequency while the asymmetries in the amplitudes of up- and down-modulated frequency harmonics become stronger by increasing the modulation frequency due to more pronounced dispersion effects which is studied in section 2 of Supplemental Material.

\subsection{Verification of Modulation-induced Phase Shift}
\begin{figure}[htbp]
\centering
{\includegraphics[width=14cm]{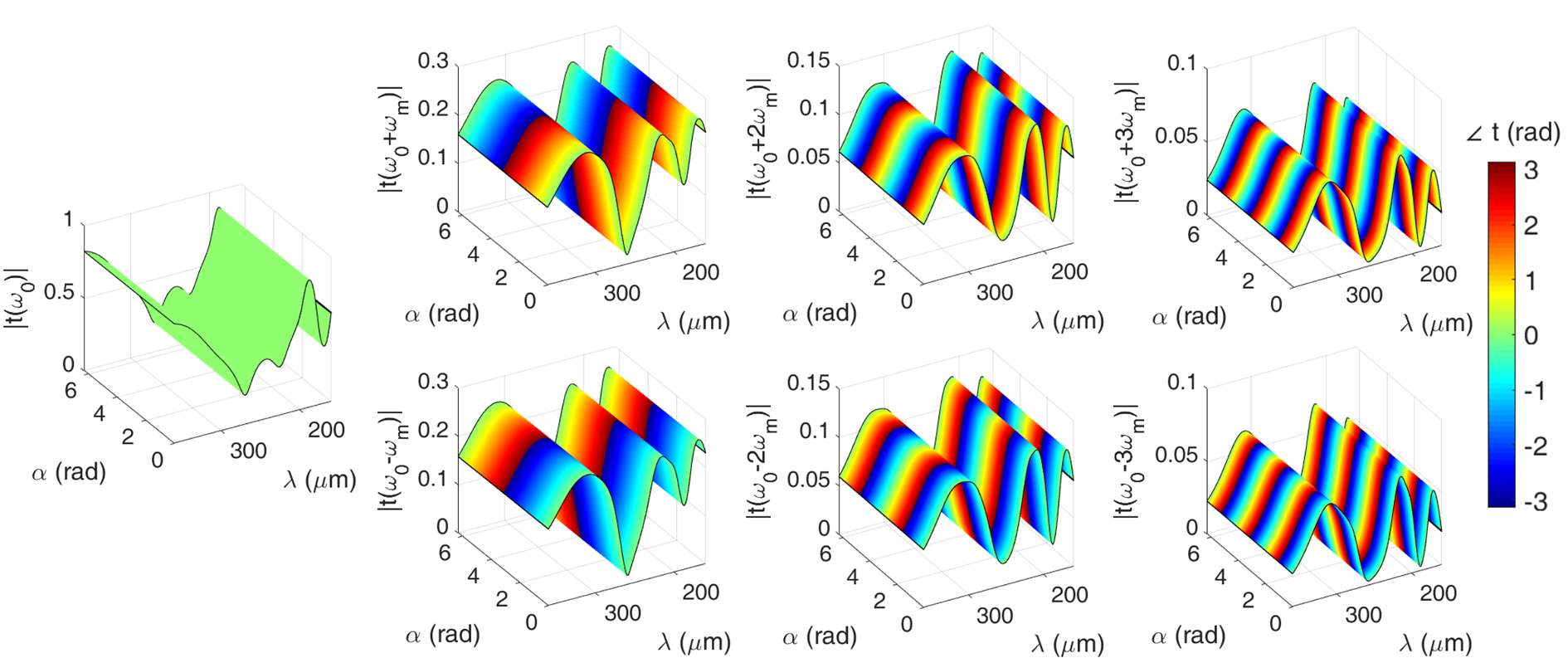}}
\caption{The calculated amplitude and phase shift of transmission coefficients corresponding to the frequency harmonics generated by the time-modulated metasurface up to the third-order, as functions of excitation wavelength and modulation phase delay for a normally incident TE-polarized plane wave. The phase shifts are indicated by pseudo-colors with the phase reference at each wavelength chosen as the phase corresponding to $\alpha=0$ to show the phase variations with respect to modulation phase delay more clearly. }
\label{fig:Fig3}
\end{figure}

In order to apply the proposed design rule and investigate the modulation-induced phase shift of light, we introduce a phase delay of $\alpha$ in the temporal modulation profile which can be implemented and electrically tuned by incorporating passive and active RF phase shifters in the biasing network of the elements. Figure \ref{fig:Fig3} demonstrates the calculated amplitude and phase shift of transmission coefficients corresponding to the generated frequency harmonics up to the third-order versus the incident wavelength and phase delay of modulation ($\alpha$), for a normally incident TE-polarized plane wave. The phase shift of frequency harmonics at each wavelength is measured with respect to their phase at the same wavelength with $\alpha=0$ and the results are shown as the pseudo-colors in the 3D plots of Fig. \ref{fig:Fig3}.  This phase reference is chosen to show the dependence of acquired phase shift by the frequency harmonics to the modulation phase delay, clearly. The results clearly verify that by introducing a modulation phase delay of $\alpha$, the transmission amplitude remains constant for all frequency harmonics while $n$-th frequency harmonic picks up a phase shift of $n\alpha$. The phase variation of frequency harmonics has the same linear profile for all wavelengths implying that the phase gradient across the metasurface is dispersionless and depends merely on the modulation phase delay which makes the metasurface functional over a broad spectral range although with varying efficiency. Moreover, the modulation induced phase shift can be seen for the higher-order multipole mode of TE$_{21}$ (EQ) which indicates that modulated-induced phase shift persists beyond the dipolar regime (TE$_{01}$ and TE$_{11}$). Similar results can be obtained for the reflection coefficient of generated frequency harmonics which are not brought here for the sake of brevity.

\begin{figure}[htbp]
\centering
{\includegraphics[width=14cm]{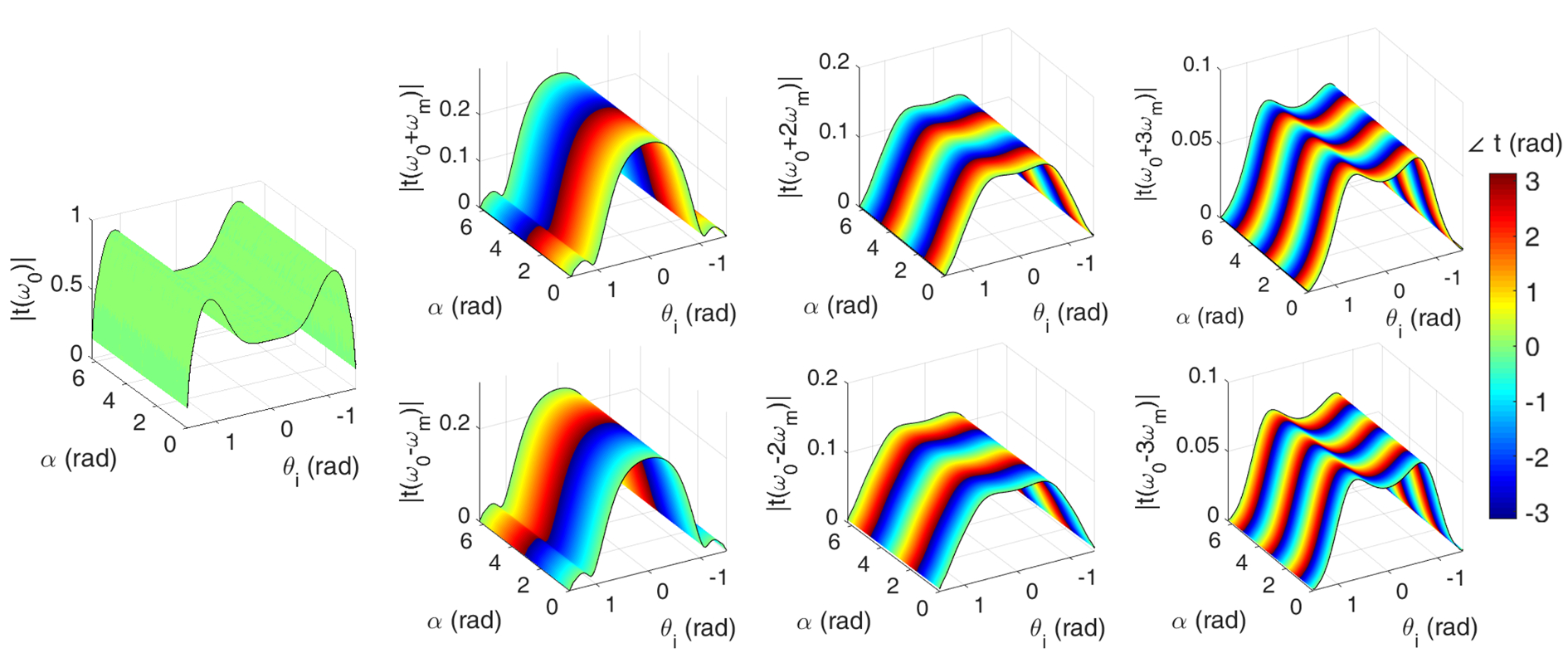}}
\caption{The calculated amplitude and phase shift of transmission coefficients corresponding to the frequency harmonics generated by the metasurface up to the third-order, as functions of incident angle and modulation phase delay for an oblique incidence of TE-polarized plane wave with an excitation wavelength of $\lambda_0=300$ $\mu$m. The phase shifts are indicated by pseudo-colors with the phase reference at each incident angle chosen as the phase corresponding to $\alpha=0$ to show the phase variations with respect to modulation phase delay more clearly. }
\label{fig:Fig4}
\end{figure}

The amplitude and phase variations with respect to the modulation phase delay for oblique incidence are also demonstrated by obtaining and plotting the transmission coefficient amplitude and phase shift of generated frequency harmonics up to the third-order as functions of modulation phase delay ($\alpha$) and incident angle ($\theta_i$) for a TE-polarized plane wave with an excitation wavelength of $\lambda_0=300$ $\mu$m in Fig. \ref{fig:Fig4}. The phase shifts are indicated by pseudo-color in 3D plots where the phase at each incident angle is measured with respect to the phase corresponding to $\alpha=0$ to show the linear phase variation profile with respect to modulation phase delay, more clearly. As it can be observed, the conversion efficiency to the frequency harmonics changes with incident angle due to the resonant characteristics of graphene-wrapped microwires while the scattering amplitude of generated frequency harmonics is independent of modulation phase delay. Furthermore, the results confirm that by introducing a modulation phase delay of $\alpha$ in an obliquely incident time-modulated metasurface, the $n$-th frequency harmonic acquires a phase shift of $n\alpha$ which implies that changing the modulation phase delay of each element can be used to realize a phase distribution across the time-modulated metasurface for all incident angles thus enabling wavefront shaping with wide angular bandwidth.   

The phase response of the metasurface to transverse magnetic (TM) polarization follows the same principle which is demonstrated in section 3 of Supplemental Material. The proposed design principle is also verified by full-wave FDTD simulations and comparisons between FDTD and semi-analytical multipole scattering simulations are included in section 4 of the Supplemental Material. 
\subsection{Frequency Conversion Efficiency}
It should be highlighted that while the modulation-induced phase is independent from the wavelength, incident angle and polarization, the frequency conversion efficiency can strongly vary as a function of these parameters depending on the geometry and resonant characteristics of the element. This implies that the metasurface can remain functional for manipulation of generated frequency harmonics, however, the power efficiency decreases away from the resonant regime. The ratio of incident power radiated in backward and forward directions at n-th frequency harmonic can be calculated using relative reflectance ($R_n$) and transmittance ($T_n$) of the periodic time-modulated metasurface. The converted power to the forward and backward propagating frequency harmonics are calculated and plotted in dB scale as functions of incident angle and wavelength of a TE-polarized plane wave illumination in Fig. 5(a) \begin{figure}[htbp]
\centering
{\includegraphics[width=0.7\textwidth]{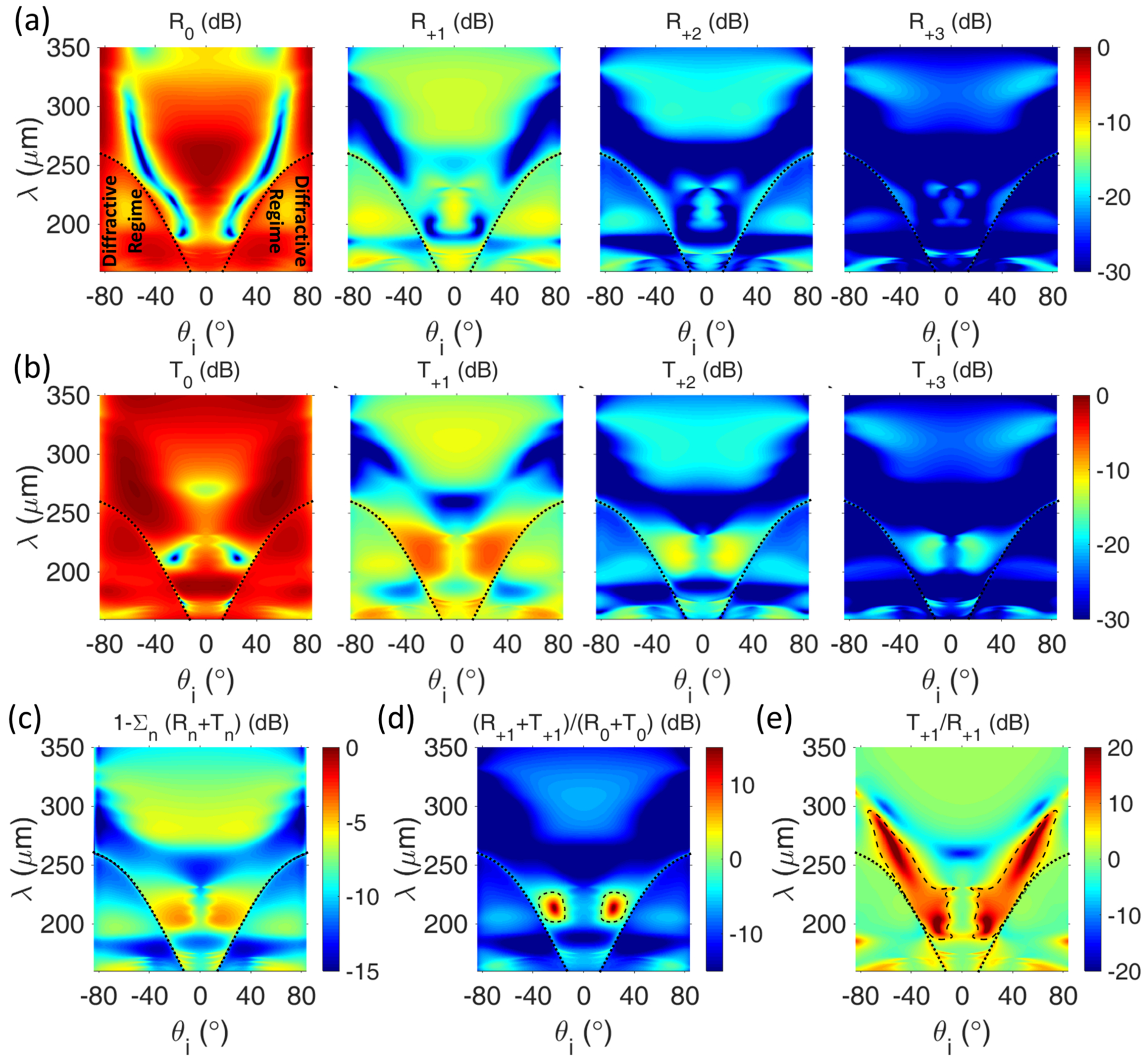}}
\caption{The calculated ratio of incident power converted to the frequency harmonics in (a) backward and (b) forward directions quantified by the relative reflectance and transmittance of time-modulated metasurface in dB scale, as functions of incident angle and wavelength of a TE-polarized plane wave illumination. The dotted lines denote the onset of diffractive regime. (c) The non-radiated ratio of incident power calculated as a function of incident wavelength and angle in dB scale, characterizing the role of absorption and material loss in the frequency conversion efficiency of metasurface. (d) The frequency conversion efficiency of the metasurface defined as the radiated power at the first-order frequency harmonics to that at the fundamental harmonic, as a function of incident angle and wavelength. The region denoted by dashed line corresponds to the maximal efficiency. (e) The directionality of the frequency conversion defined as the ratio of transmittance over the reflectance of the first-order up-modulated frequency harmonic. The dashed lines denote the region corresponding to unidirectional frequency conversion.}
\label{fig:Fig5}
\end{figure}. The results are only demonstrated for the fundamental and up-modulated frequency harmonics. The transmittance and reflectance of down-modulated frequency harmonics yield very similar results to those of up-modulated frequency harmonics which are not brought here for the sake of brevity. An important point at oblique incidence is the emergence of diffractive regime when $1\pm\sin(\theta_i)>\lambda/\Lambda$. For the unit cell under study here whose periodicity is $\Lambda=130$ $\mu$m, the higher order diffraction orders can get excited when $\lambda<260$ $\mu$m. The onset of diffractive regime is denoted by the dotted lines in color maps of Fig. 5. It should be mentioned that the plotted reflectance and transmittance of frequency harmonics correspond to the summation of the relative power of all propagating diffraction orders. The results displayed in Fig. 5 indicate the enhancement of frequency conversion efficiency in the vicinity of wavelengths corresponding to the three resonant modes (MD, ED and EQ) supported by the metasurface. Furthermore, a high frequency conversion efficiency is maintained under oblique incidence which may be attributed to the geometrical symmetry of the time-modulated scatterer and its active region (graphene wrap). The maximal conversion efficiency is limited by the absorption processes and material loss. In order to quantify the role of loss, we calculate the non-radiated ratio of incident power as $1-\sum_n(R_n+T_n)$ which is a measure of absorption. For this purpose, we have truncated the number of frequency harmonics to $\pm5$ which ensures the convergence of the result due to the negligibly small contribution of frequency harmonics with index of $|n|>5$. The calculated result for the non-radiated power is shown in dB scale as a function of incident angle and wavelength of the TE-polarized plane wave in Fig. 5(c). A considerable loss can be observed in the regions corresponding to the resonant enhancement of higher-order frequency harmonics in the account of stronger light-matter interactions in these regions which highlights the role of material loss in limiting the maximal conversion efficiency. The effect of loss can be compensated by using stronger modulation depth which is equivalent to the increment in the pumping strength of modulation signal. It should be remarked that the time-modulated metasurface is an open system in which the power of transmitted and reflected harmonics is provided in part by the external agent responsible for modulation \cite{zurita2009reflection,zurita2010resonances,taravati2017nonreciprocal2,chamanara2017optical}. Here, we have limited ourselves to the range $E_f=0-0.6$ eV for the modulation of Fermi energy level of graphene which has been shown to be accessible by electrical gating in experiments \cite{sherrott2017experimental}. Fermi energy levels as high as $E_f=1-2$ eV have been reported \cite{efetov2010controlling} in the static cases which can potentially enable higher modulation depths. A regime of operation which deserves our attention is the oblique incidence regime ($\theta_i\approx23^{\circ}$) around the incident wavelength of $\lambda=214$ $\mu$m which corresponds to the maximal frequency conversion in the forward direction and maximal absorption. In order to illustrate this more clearly, we define the frequency conversion efficiency as the ratio of the total radiated power at the first-order frequency harmonic to that at the fundamental frequency harmonic ($\eta=\frac{T_{+1}+R_{+1}}{T_0+R_0}$) and the calculated result is plotted in Fig. 5(d) as a function of incident wavelength and angle. Despite the enhancement of the conversion efficiency around the resonant modes, most of the radiated power still resides at the fundamental frequency harmonic except for the critical region denoted by the dashed lines where the transmittance and reflectance of fundamental frequency harmonic reach their minimum values while the transmittance of higher-order frequency harmonics is significantly enhanced. Furthermore, in this region the transmittance of higher-order frequency harmonics is significantly dominant over their reflectance. This can be verified by defining directionality of frequency conversion as the transmittance of first-order up-modulated frequency harmonic over its reflectance ($D=\frac{T_{+1}}{R_{+1}}$) which is calculated and plotted in Fig. 5(e) as a function of incident wavelength and angle. The frequency conversion is observed to be mostly bidirectional except in the narrow regime of oblique incidence denoted by the dashed line. The maximal frequency conversion efficiency and unidirectionality of frequency harmonic generation are manifestations of a dynamic Huygens regime with overlapped electric and magnetic responses as established recently in \cite{liu2018huygens}. Further details and discussions related to this regime and its physical origin are brought in the section 5 of the Supplemental Material. It should be noted that despite the high conversion efficiency and directionality afforded in the Huygens regime, it is strongly dependent on the incident angle and wavelength \cite{arslan2017angle,paniagua2016generalized}. Higher conversion efficiencies may be maintained over the spectral and angular regime of operation using elements with more exotic geometries exhibiting multimodal directionality \cite{yang2017multimode} or using multigate active elements \cite{lee2017multiple,forouzmand2018tunable,kafaie2018dual} with cascaded time-modulated layers which can also potentially serve as isolators by engineering the bandgaps to realize an asymmetric dispersion \cite{chamanara2017optical,taravati2017nonreciprocal2}.

\section{Wavefront Engineering of Generated Frequency Harmonics}
\begin{figure}[htbp]
\centering
{\includegraphics[width=10cm]{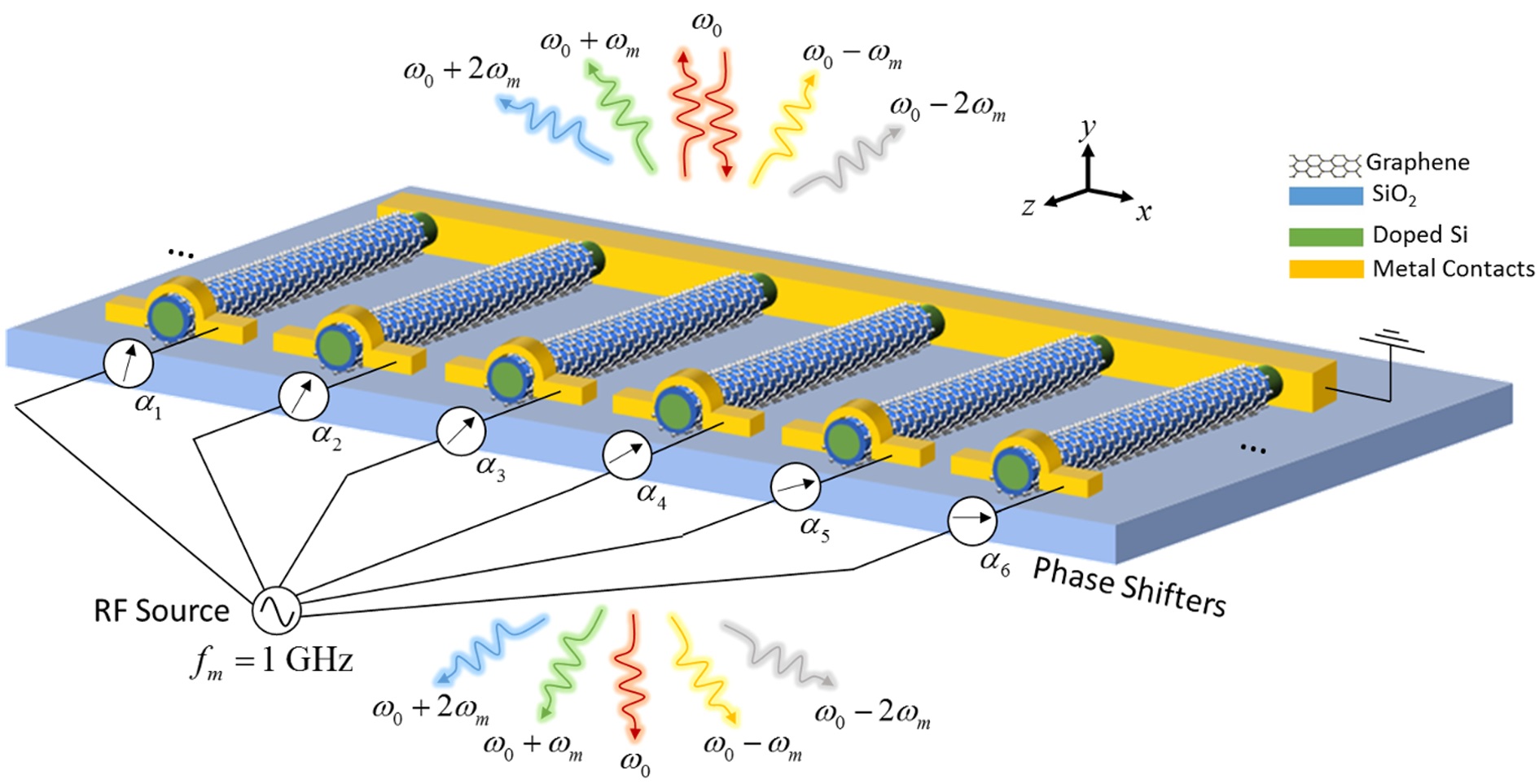}}
\caption{The schematic of a phased time-modulated array consisted of graphene-wrapped microwires for wavefront engineering of generated frequency harmonics.}
\label{fig:Fig6}
\end{figure}

In this section, we demonstrate the utility of the proposed design rule and time-modulated metasurface for wavefront engineering of the generated frequency harmonics. For this purpose, we use a synthesis method based on generalized Snell's laws which intuitively allows for arbitrary control of reflected and transmitted waves through local phase shifts at the metasurface. It should be mentioned that this synthesize method suffers in terms of power efficiency due to spurious scattering and it has been recently found that the ideal required phase shift differs from the prescription of the generalized Snell's law \cite{estakhri2016wave,asadchy2016perfect,epstein2016synthesis}. However, due to the $2\pi$ phase span provided by the proposed design paradigm, it can be used flexibly in more sophisticated synthesis methods to improve the efficiency of functionality. In the following, we consider a normally incident plane wave with TE polarization and wavelength of $\lambda_0=300$ $\mu$m and investigate beam steering and focusing of generated frequency harmonics as classical examples of wavefront engineering. Due to the conjugate phase shift of the up- and down- modulated harmonics and the difference between the acquired phase shifts by different harmonic orders, several exotic behaviors can arise which are comprehensively studied and discussed. For this purpose, we model a large finite array of 101 microwires (with overall length of $\approx 43.77\lambda_0$) which is geometrically fixed but the modulation phase delay can be changed electrically through independent addressing and biasing each microwire to realize the required phase distribution for attaining a certain functionality at higher-order frequency harmonics. As such, the metasurface is modulated in both space and time (exhibits a space-time gradient) and allows for realization of spatially varying phase discontinuities at higher-order frequency harmonics. Figure \ref{fig:Fig6} depicts the schematic of such a phased time-modulated array. The required phase shifts in modulation can be provided by an array of passive or active phase shifters in RF modulation circuit \cite{firouzjaei2010mm}.

\subsection{Steering of Generated Frequency Harmonics}
As the first application, we aim at steering the generated frequency harmonics toward specific directions. To this purpose, the metasurface should provide a linear phase gradient along one direction at the desired frequency harmonic. The bending angle of n-th frequency harmonic ($\theta_n$) will be defined by the phase change between neighbor elements at n-th frequency harmonic ($\Delta\phi_n$), according to the generalized Snell's law for anomalous refraction as \cite{yu2014flat}:
\begin{equation}
\Delta\phi_n=-k_n\sin(\theta_n)\Lambda
\end{equation}
where $k_n=(\omega_0+n\omega_m)/c$ is the wavenumber corresponding to the n-th frequency harmonic and $\Lambda$ is the periodicity. For a metasurface in which the unit cells are not deeply subwavelength or the phase variation between the neighbor elements is very rapid such that $|\Delta\phi_n/k_n\Lambda|>1$, the anomalous diffraction may also occur, whose contribution can be taken into account using the generalized diffraction equation as \cite{huang2012dispersionless}:
\begin{equation}
\Delta\phi_n+2m\pi=-k_n\sin(\theta_{n,m})\Lambda
\end{equation}
where $m$ is the diffraction order. According to the theory and the results presented in previous sections, the linear phase gradient can be achieved by varying the modulation phase delay ($\alpha$) linearly across the metasurface. A linear profile for the modulation phase delay given by the progressive delay of $\Delta\alpha$ translates to a linear phase gradient profile for all higher-order frequency harmonics, where the phase gradient of n-th frequency harmonic is defined by the progressive delay of $\Delta\phi_n=n\Delta\alpha$. This will lead to spatial decomposition of generated frequency harmonics as they will simultaneously bend toward well-separated angles defined by $\theta_{n,m}=-\arcsin((n\Delta\alpha+2m\pi)/\Lambda k_n)$. In particular, up- and down-modulated frequency harmonics will experience opposite phase gradients and will anomalously bend into opposite directions. In the case of modulation frequency being small compared to the optical frequency, $\frac{k_{+n}-k_{-n}}{k_0}\ll 1$ which results into symmetric bending angles for up- and down- modulated frequency harmonics $\theta_{+n,+m}\approx-\theta_{-n,-m}$.

\begin{figure}[htbp]
\centering
{\includegraphics[width=12cm]{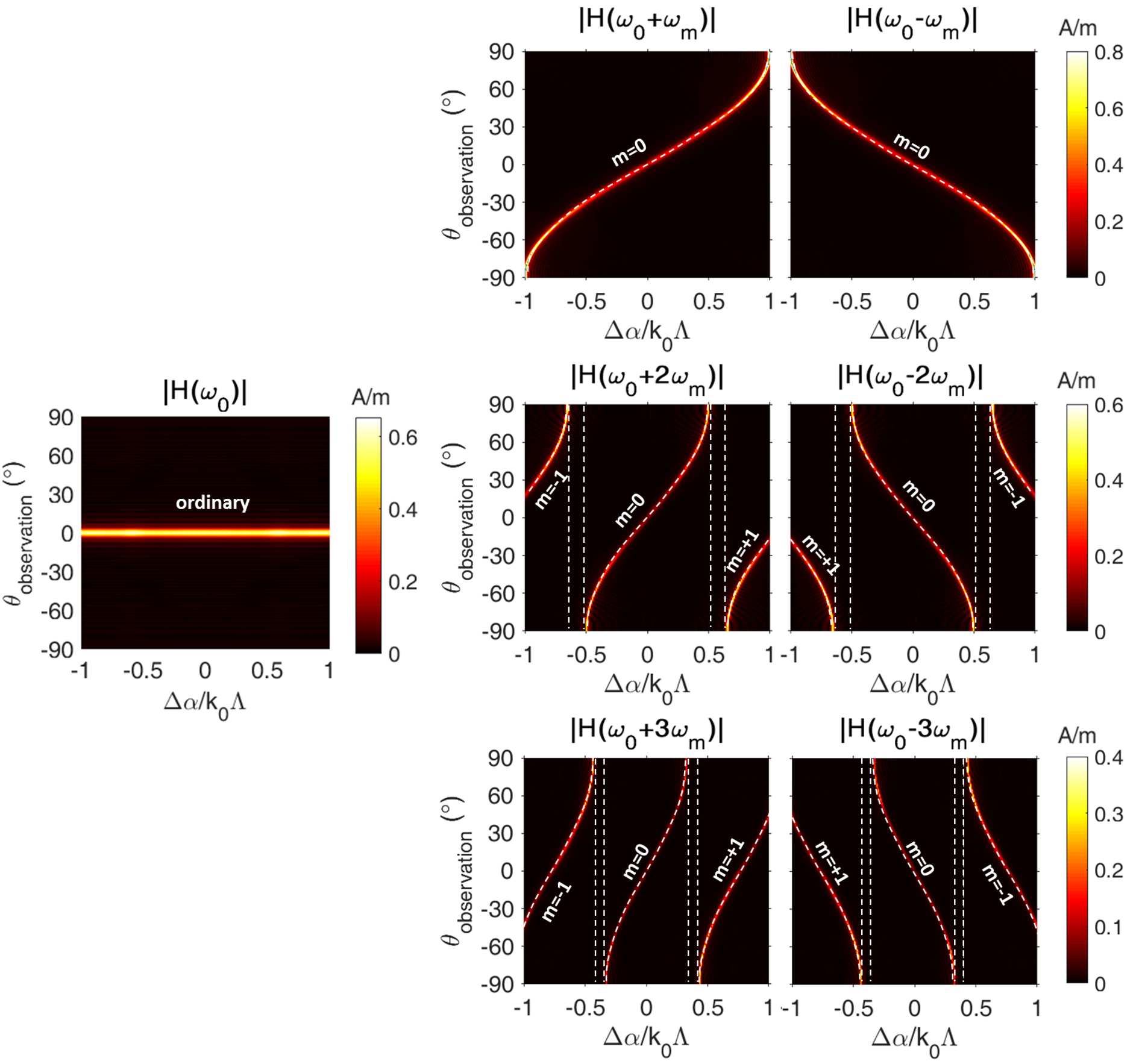}}
\caption{The amplitude of magnetic field at generated frequency harmonics by the time-modulated metasurface with a linear modulation phase delay profile, calculated as a function of observation angle in the farfield region and the progressive modulation phase delay ($\Delta\alpha$), for a normal incidence of TE-polarized plane wave with $\lambda_0=300$ $\mu$m. $m$ denotes diffraction order of the branches.}
\label{fig:Fig7}
\end{figure}

Figure \ref{fig:Fig7} shows the farfield patterns of magnetic field at different frequency harmonics as functions of progressive modulation phase delay ($\Delta\alpha$) which are obtained by evaluating the amplitude scattered magnetic fields from the finite time-modulated metasurface in the backward direction (reflected fields) at different observation angles at a radial distance of $100L$ from the center of the metasurface with $L$ being the overall length of the metasurface. As it can be clearly observed from the results, the bending angle of fundamental frequency harmonic is governed by the ordinary refraction as it does not experience any phase gradient while all the higher-order harmonics undergo anomalous refraction/diffraction due to the modulation-induced phase gradient. By increasing $|\Delta\alpha|$ from $0^{\circ}$, the bending angle of n-th frequency harmonic increases with a dependency of $\theta_n=-\arcsin(n\Delta\alpha/\Lambda k_n)$ and asymptotically approaches end-fire directions $|\theta_n|=90^{\circ}$ as $|n\Delta\alpha/\Lambda k_n|$ reaches 1. Increasing $\Delta\alpha$ beyond this point, the sign of bending angle is flipped and the anomalous diffraction orders $m=\pm 1$ start to appear due to rapid phase variation between neighbor elements, consistent with the generalized diffraction (Eq. (20)). Interestingly, there exist narrow shadow regions between the branches of $m=0$ and $m=\pm 1$ anomalous diffraction orders with no beams. In these regions, the generated frequency harmonics are coupled to the evanescent modes and travel along the metasurface as guided modes with no radiation to free-space. The results clearly illustrate that the afforded $2\pi$ span in the phase shift of generated frequency harmonics enables an electrical beam-scanning with $180^{\circ}$ angle-of-view. Such all-angle scanning performance cannot be achieved in most of quasi-static active metasurfaces due to the limitations in the tunability range and material loss which limit the spectral shift of the resonance and depress the resonant phase agility, respectively. The converted power into the frequency harmonics are reported in section 7 of Supplemental Material.

\begin{figure}[htbp]
\centering
{\includegraphics[width=\textwidth]{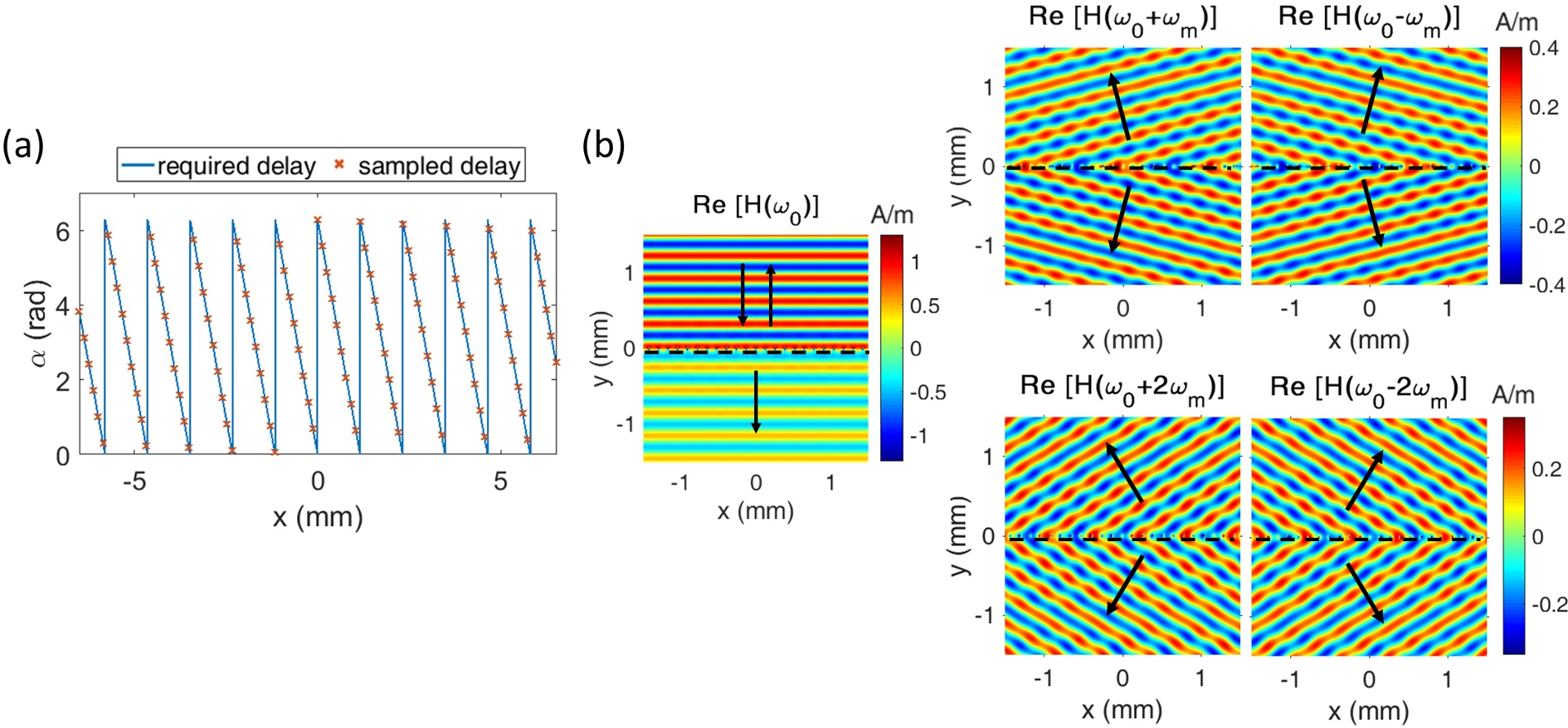}}
\caption{(a) The linear modulation phase delay profile across the time-modulated metasurface with a gradient given by $\Delta\alpha=-k_1\Lambda\sin(15^{\circ})$. (b) The simulated wavefronts of the magnetic field corresponding to the generated frequency harmonics by the metasurface under illumination of a normally incident TE-polarized plane wave with $\lambda_0=300$ $\mu$m.}
\label{fig:Fig8}
\end{figure}

In order to evaluate the nearfield performance of the time-modulated metasurface in steering the frequency harmonics, we set $\Delta\alpha/k_1\Lambda=-\sin(15^{\circ})$ which is chosen to bend the first up-modulated frequency harmonic toward the angle of $\theta_1=15^{\circ}$. The modulation phase delay profile across the metasurface and the magnetic field wavefronts of generated frequency harmonics up to the second order are demonstrated in Figs. \ref{fig:Fig8}(a) and (b), respectively. The metasurface is quasi-transparent for the fundamental frequency harmonic as it does not experience any phase gradient. The first up-modulated harmonic is bent toward the desired angle of $\theta_1=15^{\circ}$ in both forward and backward scattering planes. The down-modulated harmonic bends into the opposite direction by virtue of experiencing opposite phase gradient. Moreover, the second-order generated frequency harmonics bends toward $\theta_{\pm2}=\pm\arcsin(2k_1/k_{\pm2}\sin(15^{\circ}))\approx\pm31^{\circ}$. The bending angles of generated frequency harmonics can be tailored at will, by electrically adjusting the modulation phase delay across the geometrically-fixed time-modulated metasurface.  

It should be emphasized that the generation of higher-order frequency harmonics in both forward and backward directions is due to the geometrical symmetry of the modulated scatterers and non-overlapping electric and magnetic resonances in this operation regime which results into a bidirectional radiation pattern whose symmetry is only slightly broken by the presence of low index dielectric substrate. This is also the case in single-layer nonlinear metasurfaces \cite{keren2015nonlinear} and linear plasmonic metasurfaces shaping the cross-polarized light \cite{shi2014coherent,ni2013ultra,chen2012dual,tymchenko2016advanced} which are bound to re-radiate and scatter the light on the opposite sides. Such bidirectional metasurfaces can act as reflect-transmit-arrays which are promising candidates for wireless communication applications \cite{yang2018design}. However, a unidirectional generation of harmonics can increase the conversion efficiency and may be more desirable in some scenarios. This can be achieved in the transmission mode through dynamic Huygens regime of time-modulated metasurfaces employing both electric and magnetic responses \cite{liu2018huygens}. A unidirectional reflect-array can also be obtained by simply placing a back-mirror to block the transmission. In such cases, the design rule persists while the conversion efficiency is also increased. Further results and discussions on the transmittive and reflective time-modulated metasurfaces are provided in sections 5 and 6 of the Supplemental Material.

\subsection{Focusing of Generated Frequency Harmonics}
A great opportunity offered by electrically tunable metasurfaces is achieving multiple functionalities through a single platform. Here, we demonstrate focusing of generated frequency harmonics using the implemented time-modulated metasurface as another application. To this purpose, the phase gradient across the metasurface should follow a quadratic profile, as \cite{yu2014flat}:
\begin{equation}
\phi_n(x)=-k_n(\sqrt[]{x^2+F_n^2}-F_n)
\end{equation}
where $F_n$ is the focal distance of n-th frequency harmonic. The quadratic phase profile can be realized by changing the modulation phase delay across the metasurface, correspondingly. Unlike the bending application where a linear modulation phase delay profile translates into a linear phase gradient profile for all frequency harmonics and leads to simultaneous bending of all harmonics, in this case the focusing requirement for different frequency harmonics is different as a linear scaling of the quadratic phase profile does not correspond to a focusing phase profile.

\begin{figure}[htbp]
\centering
{\includegraphics[width=\textwidth]{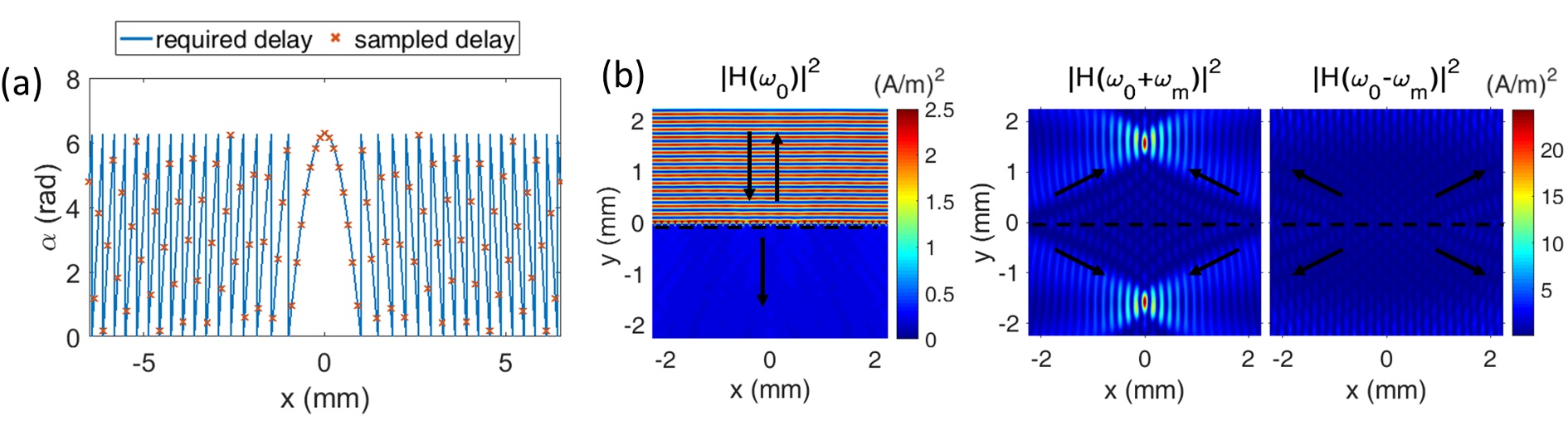}}
\caption{(a) The required quadratic modulation phase delay profile across the time-modulated metasurface to focus the first up-modulated harmonic due to a normal incidence with $\lambda_0=300$ $\mu$m at a focal distance of $F_1=1.5$ mm. (b) The simulated magnetic field intensity corresponding to the fundamental and first-order frequency harmonics generated by the metasurface.}
\label{fig:Fig9}
\end{figure}

Here, we choose the modulation phase delays in accordance to the focusing requirement of the first up-modulated harmonic at a focal distance of $F_1=5\lambda_0=1.5$ mm. The required modulation phase delay profile across the metasurface and the magnetic field intensities of the fundamental and first generated frequency harmonics are plotted in Figs. \ref{fig:Fig9}(a) and (b), respectively. Similar to the previous case, the fundamental frequency does not experience any phase gradient and is partially reflected and transmitted. The intensity of the first up-modulated harmonic indicates nearly perfect focusing at the desired focal distance from the metasurface plane in both forward and backward scattering planes. Moreover, due to opposite phase gradient experience by the down-modulated harmonic, it is strongly diffused and diverged away from the center metasurface creating a shadow region at the center. As such, this time-modulated lens has dual-polarity functionality as it is convex for up-modulated frequency harmonic and concave for down-modulated frequency harmonic.

\section{Spectral Performance}

As established by formulation and verified from the results shown in Fig. \ref{fig:Fig3}, the modulation-induced phase shift is dispersionless. This implies that the metasurface can remain functional over a broadband spectrum, although with different efficiencies due to the dependence of frequency conversion efficiency to the resonant characteristics of the element. This offers a great advantage compared to quasi-static tunable metasurfaces in which the resonant phase agility is limited to an extremely narrow bandwidth outside of which the metasurface loses its functionality \cite{sherrott2017experimental,huang2016gate,park2016dynamic}. In this section, we study the spectral performance of the functional time-modulated metasurface in steering and focusing of frequency harmonics to verify their beam shaping capabilities in a broadband spectrum. 

\begin{figure}[htbp]
\centering
{\includegraphics[width=12cm]{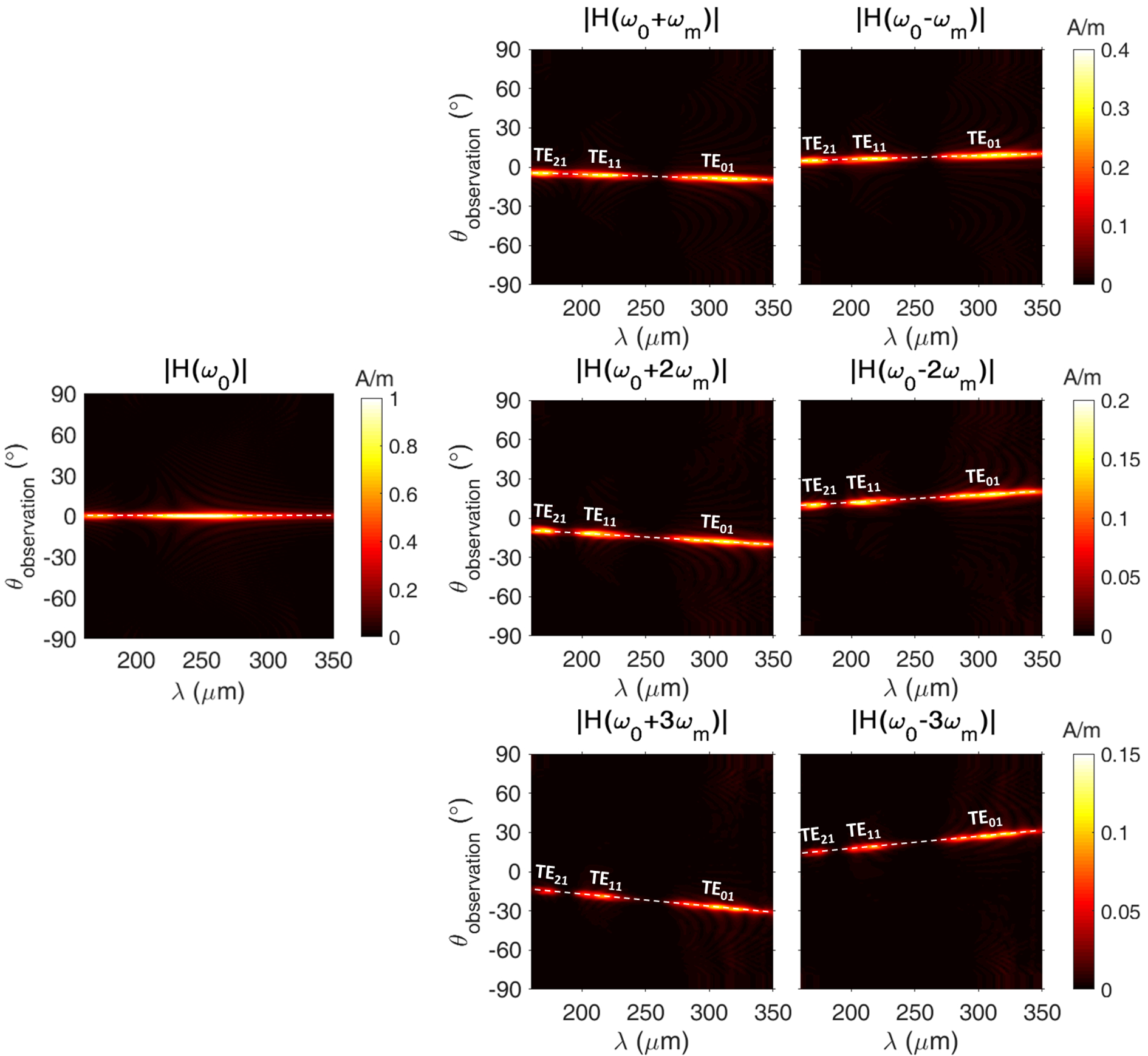}}
\caption{The magnetic field amplitude of generated frequency harmonics by the time-modulated metasurface with a linear modulation phase delay profile defined by $\Delta\alpha=2\pi\frac{\Lambda}{350\mu \text{m}}\sin(10^{\circ})$, calculated as a function of observation angle in the farfield region and incident wavelength, for a normal incidence of TE-polarized plane wave.}
\label{fig:Fig10}
\end{figure}

First, we consider the time-modulated metasurface having a fixed linear modulation phase delay profile with a progressive delay of $\Delta\alpha=2\pi\frac{\Lambda}{350 \mu \text{m}}\sin(10^{\circ})$, illuminated by a normally incident TE-polarized plane wave and study its broadband performance in beam steering. The farfield patterns of magnetic field at different frequency harmonics versus the incident wavelength are demonstrated in Fig. \ref{fig:Fig10} by calculating the amplitude of magnetic field at different observation angles in farfield region. In the account of dispersionless modulation-induced phase shift and fixed modulation profile, the phase gradient is constant and has a linear profile across the metasurface for each frequency harmonic at all wavelengths and the anomalous bending angle varies with the wavelength according to Eq. (19). Decreasing the wavelength, the bending angle of higher-order frequency harmonics approaches broadside direction ($\theta_n=0^{\circ}$), and increasing the wavelength causes the beam to move toward end-fire directions ($|\theta_n|=90^{\circ}$). It should be noted that while the functionality of the metasurface is preserved in the broadband spectrum, the frequency conversion efficiency is strongly dependent on the wavelength. In particular, three resonant peaks can be observed in the spectral amplitude of the frequency harmonics which correspond to the resonant enhancement of frequency conversion at TE$_{01}$, TE$_{11}$ and TE$_{21}$ modes. 

\begin{figure}[htbp]
\centering
{\includegraphics[width=14cm]{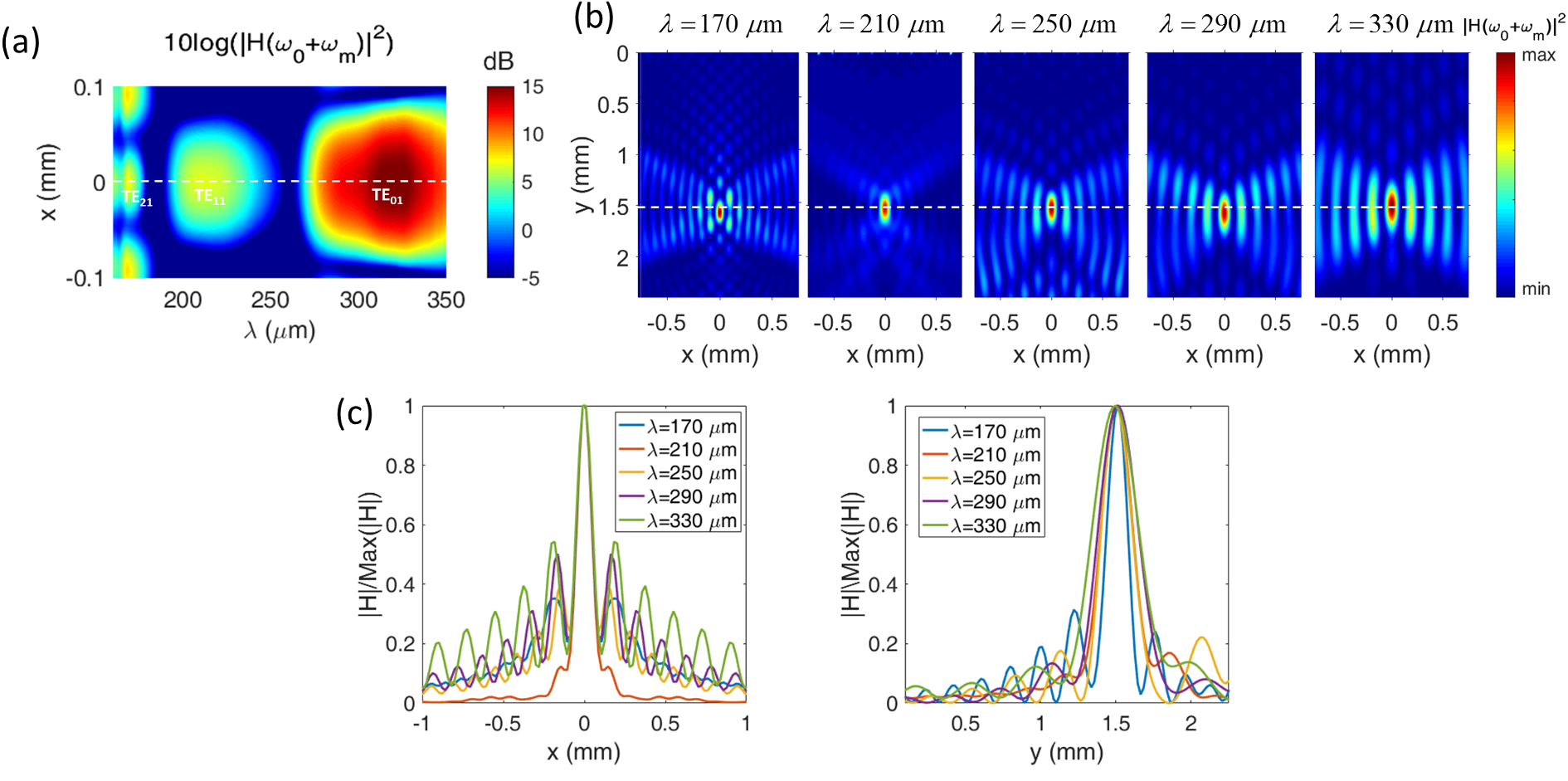}}
\caption{(a) The magnetic field intensity corresponding to the first up-modulated frequency harmonic generated by the time-modulated metasurface under illumination of a normally incident TE-polarized plane wave at the focal distance of $F_1=1.5$ mm, calculated as a function of incident wavelength and transverse position. The modulation phase delay is wavelength-dependent and is changing quadratically across the metasurface to focus the first up-modulated harmonic at a focal distance of $F_1=1.5$ mm. (b) The simulated nearfield distributions corresponding to the normalized magnetic field intensity at the first up-modulated frequency harmonic generated by the time-modulated metasurface for different incident wavelengths. (c) The normalized magnetic field intensities corresponding to different incident wavelengths calculated at the focal plane along x- direction and at the center along y-direction for clear demonstration of width and depth of focusing. }
\label{fig:Fig11}
\end{figure}

Next, we examine the spectral performance of the time-modulated metasurface in focusing of generated frequency harmonics. It should be noted that unlike steering, the focusing requirement at different wavelengths is different and in order to simultaneously achieve an aberration-free focusing at all wavelengths, the metasurface should possess a linear phase response over the operating spectrum. This is while the modulation induced phase shift in time-modulated metasurfaces is dispersionless (constant phase response over the operating spectrum) which hinders the possibility of simultaneous aberration-free focusing at different wavelengths. As such, the focusing of a polychromatic excitation cannot be achieved. However, the real-time tunability of the dynamic metasurface allows for adjusting the modulation phase delay adaptively, according to the incident wavelength to compensate the chromatic aberrations and achieve monochromatic focusing at different wavelengths. As such, we consider a wavelength-dependent quadratic modulation phase delay for the metasurface in accordance to the focusing requirement of the first up-modulated frequency harmonic at a focal distance of $F_1=1.5$ mm, under illumination of a normally incident TE-polarized plane wave and investigate the focusing performance at different wavelengths. Fixing the focal distance and size of the array, the numerical aperture (NA) of the lens remains constant at all wavelengths. Figure \ref{fig:Fig11}(a) 
represents the magnetic field intensity of first up-modulated frequency harmonic in logarithmic scale vs position and wavelength at the focal plane $F_1=1.5$ mm. The nearfield results corresponding to the normalized magnetic field intensity of the first up-modulated frequency harmonic for different excitation wavelengths are also shown in Fig. \ref{fig:Fig11}(b).
An almost perfect focusing is achieved at the desired location for all the incident wavelengths which verifies the functionality of the metasurface at the operating spectrum. In order to clearly demonstrate the width and depth of focusing, the normalized magnetic field intensity is plotted in Fig. 11(c) at different wavelengths, in the focal plane along the transverse $x$-direction and at the center along $y$-direction. The size of focal spot is determined by the diffraction limit and becomes larger by increasing the excitation wavelength. It should be noted that while the numerical aperture of the lens is fixed at all wavelengths, the effective of size of the metasurface in terms of wavelength decreases and the focal spot gets closer to the nearfield region by increment of the wavelength which can explain the appearance of speckle noises in the focal plane. The frequency conversion efficiency is dependent on the incident wavelength and the spectral intensity of magnetic field corresponding to the first up-modulated harmonic exhibits three peaks at the resonant wavelengths. 

The spectral performance of the finite metasurfaces studied in this section in terms of power conversion to the frequency harmonics is discussed in the section 7 of Supplemental Material.

\section{Angular Performance}
Another great benefit of the modulation-induced phase shift is its independence from the incident angle (as established by the formulation and implied by the results in Fig. 4). This allows the metasurface to remain functional for a wide range of incident angles. However, similar to the spectral behavior of functional time-modulated metasurfaces, the efficiency will be dependent on the incident angle in the account of changes in frequency conversion efficiency. The constant phase response of functional time-modulated metasurfaces versus incident angle can increase the acceptance angle which is severely limited in quasi-static tunable metasurfaces due to the strong resonant dispersion.    

In this section, we study the steering and focusing performance of the functional time-modulated metasurface under oblique incidence. It should be noted that when the metasurface is incident by an oblique plane wave, a phase gradient due to the tangential momentum of the incidence is added to the phase gradient achieved by modulation phase delay. As such, the acquired phase shift by the n-th frequency harmonic generated by the time-modulated metasurface upon illumination by an oblique plane wave with incident angle of $\theta_{\text{incident}}$ and angular frequency of $\omega_0$ is equal to $n\alpha-k_0\sin(\theta_{\text{incident}})x$ with $k_0=\omega_0/c$ and $x$ being the transverse direction of the metasurface. 

\begin{figure}[htbp]
\centering
{\includegraphics[width=12cm]{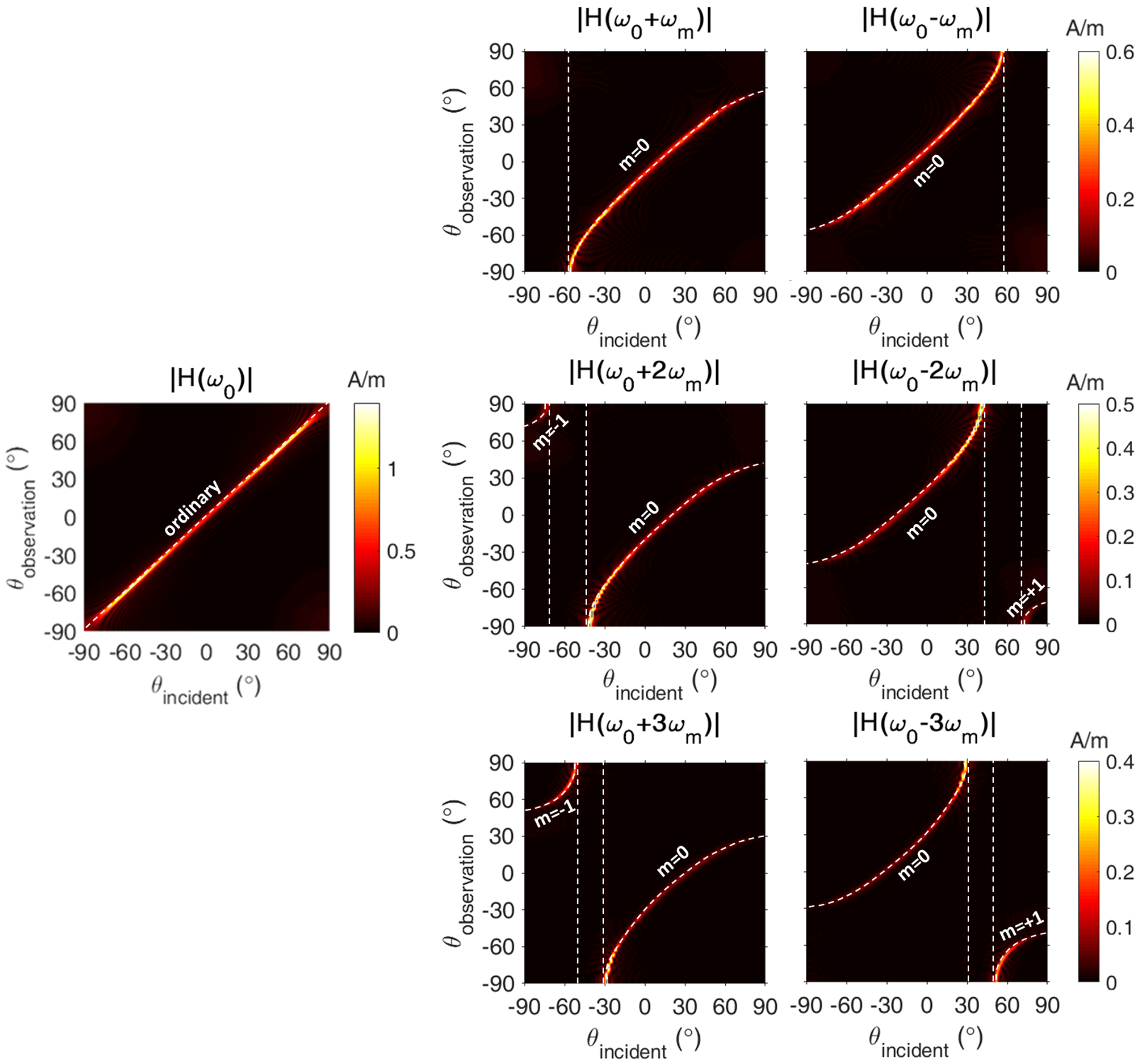}}
\caption{(a) The magnetic field amplitude at generated frequency harmonics by the time-modulated metasurface with a linear modulation phase delay profile defined by $\Delta\alpha=k_0\Lambda\sin(10^{\circ})$, calculated as a function of observation angle in the farfield region and incident angle, for an oblique incidence of TE-polarized plane wave with $\lambda_0=300$ $\mu$m.}
\label{fig:Fig12}
\end{figure}

The governing equation on anomalous refraction/diffraction for a metasurface under oblique incidence at an angle of $\theta_{\text{incident}}$, can be written as \cite{yu2014flat,huang2012dispersionless}:
\begin{equation}
\Delta\phi_n+2m\pi-k_0\sin(\theta_{\text{incident}})\Lambda=-k_n\sin(\theta_{n,m})\Lambda
\end{equation}
The addition of the off-set term due to tangential momentum of obliquely incident light breaks the symmetry between up- and down-modulated frequency harmonics in the transverse direction with respect to the surface normal. As such, the bending angles of the up- and down-modulated harmonics will not be opposite of each other anymore. Figure \ref{fig:Fig12} depcits the farfield patterns of magnetic field at different frequency harmonics versus incident angle corresponding to the time-modulated metasurface
having a fixed modulation phase delay profile with a progressive delay of $\Delta\alpha=k_0\Lambda\sin(10^{\circ})$, illuminated by an oblique TE-polarized plane wave with a wavelength of $\lambda_0=$ 300 $\mu$m. Similar to the case of normal incidence, the fundamental frequency harmonic is governed by the ordinary refraction, while all higher-order frequency harmonics exhibit anomalous bending. The results clearly indicate the strong asymmetry of up- and down-modulated frequency harmonics for oblique incident angles ($\theta_{\text{incident}}\neq 0^{\circ}$) with respect to the surface normal ($\theta_{\text{observation}}=0^{\circ}$). Increasing (decreasing) the incident angle, the down- (up-) modulated frequency harmonics tilt toward end-fire directions with a monotonic increment in the efficiency. For large incident angles, the anomalous diffraction orders start to appear on opposite sides which are separated from the anomalous refraction branch with the shadow regions corresponding to the excitation of guided modes along the metasurface.

The nearfield performance of the metasurface in steering the frequency harmonics under oblique incidence is investigated by considering the time-modulated metasurface with $\Delta\alpha=k_0\Lambda\sin(10^{\circ})$ under illumination of an obliquely incident TE-polarized plane wave at an angle of $\theta_{\text{incident}}=-40^{\circ}$ and a wavelength of $\lambda_0=300$ $\mu$m. The magnetic field wavefronts of generated frequency harmonics in Fig. 13 \begin{figure}[htbp]
\centering
{\includegraphics[width=10cm]{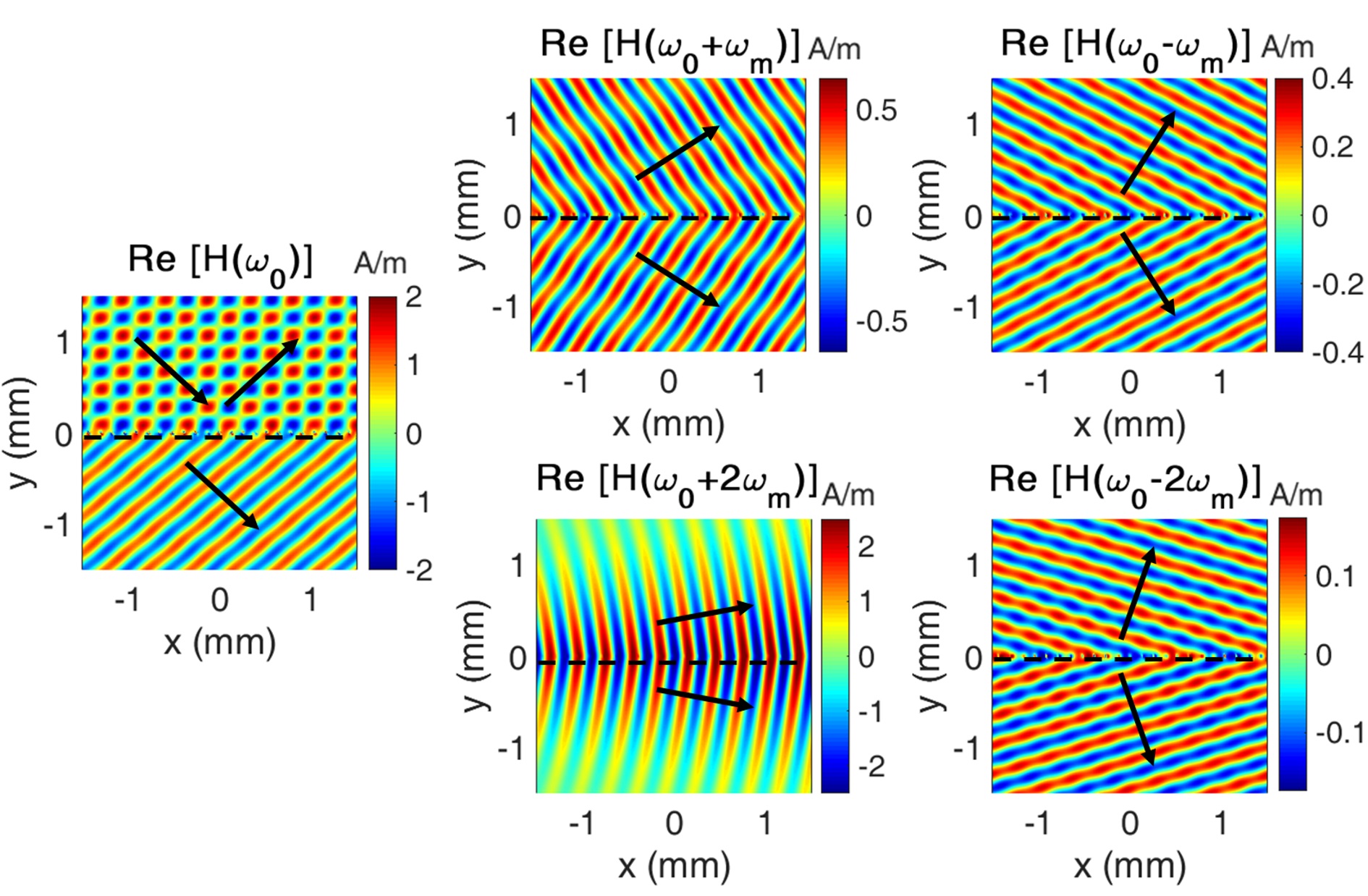}}
\caption{The simulated wavefronts of magnetic field corresponding to the generated frequency harmonics by the time-modulated metasurface with a linear modulation phase delay profile defined by $\Delta\alpha=k_0\Lambda\sin(10^{\circ})$ under illumination of an obliquely incident TE-polarized plane wave at angle of $\theta_{\text{incident}}=40^{\circ}$ with $\lambda_0=300$ $\mu$m.}
\label{fig:Fig13}
\end{figure} show the ordinary refraction/reflection of fundamental harmonic and anomalous bending of up- and down-modulated toward different angles which are not symmetric with respect to the surface normal. The bending angles are calculated using Eq. (22) as $\theta_{+1}\approx-55^{\circ}$, $\theta_{-1}\approx-28^{\circ}$, $\theta_{+2}\approx-81^{\circ}$ and $\theta_{-2}\approx-17^{\circ}$. The nearfield results are in perfect agreement with the theoretical predictions.

\begin{figure}[htbp]
\centering
{\includegraphics[width=14cm]{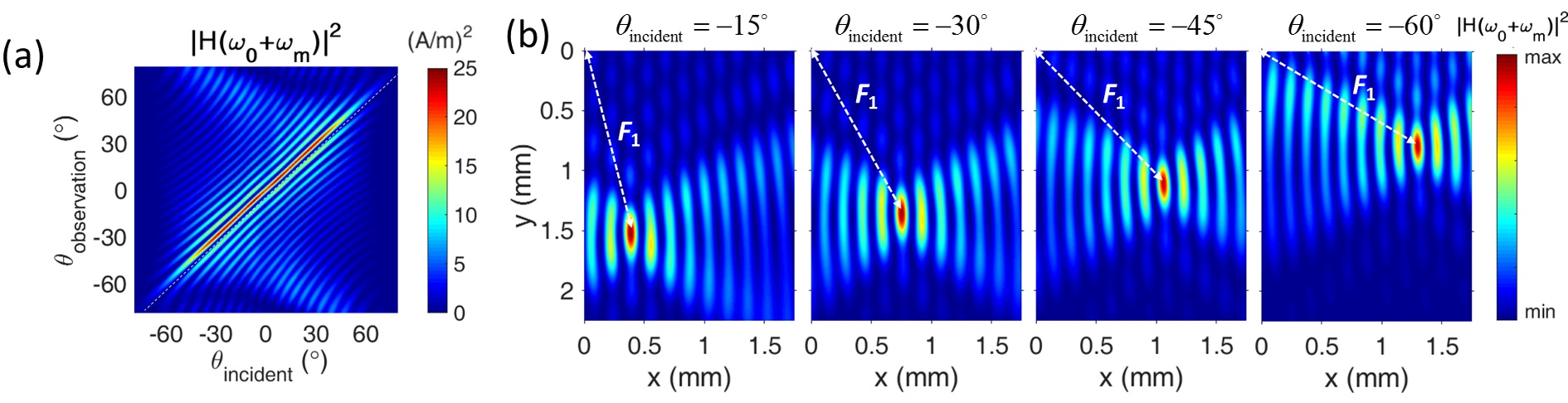}}
\caption{(a) The magnetic field intensity of the first up-modulated frequency harmonic generated by the time-modulated metasurface under illumination of an obliquely incident TE-polarized plane wave with $\lambda_0=300$ $\mu$m at the focal distance of $F_1=1.5$ mm, calculated as a function of incident angle and observation angle. The modulation phase delay is angle-dependent and is changing quadratically across the metasurface to focus the first up-modulated harmonic at a focal distance of $F_1=1.5$ mm in an off-axis direction set by the incident angle. (b) The simulated nearfield distributions corresponding to the normalized magnetic field intensity of the first up-modulated frequency harmonic generated by the time-modulated metasurface for different incident angles.}
\label{fig:Fig14}
\end{figure}

Next, we evaluate the angular performance of the metasurface in focusing generated frequency harmonics. In order to achieve an aberration-free focusing for different incident angles, the additional transverse phase gradient due to the oblique incidence should be canceled out which makes the required phase profile angle-dependent. Moreover, the position of focal point in a perfect focusing lens is a function of incident angle as it should focus the light in an off-axis direction under oblique incidence \cite{deng2016wide,kalvach2016aberration,arbabi2016miniature}. The angle-dependent phase profile of the focusing metasurface can be expressed as \cite{deng2016wide}:
\begin{equation}
\phi_n(x)=-k_n(\sqrt[]{(x+F_n\sin(\theta_{inc}))^2+F_n^2\cos^2(\theta_{inc})}-F_n)+k_0x\sin(\theta_{\text{incident}})
\end{equation}
Similar to the spectral performance, simultaneous focusing of different incident angles is not possible due to the constant phase response of time-modulated metasurface with respect to the incident angle. However, the real-time tunability of the active time-modulated metasurface enables adaptive adjusting of the phase profile, according to the incident angle to compensate the aberrations resulting from oblique incidence and to yield a wide-angle focusing response. We consider an angle-dependent quadratic phase profile according to Eq. (22) for focusing the first up-modulated frequency harmonic at an off-axis focal distance of $F_1=1.5$ mm. The metasurface is obliquely incident by a TE-polarized plane wave with a wavelength of $\lambda_0=$300 $\mu$m. Figure \ref{fig:Fig14}(a) demonstrates the magnetic field intensity of the first up-modulated frequency harmonic at the focal distance of $F_1=1.5$ mm vs observation angle and incident angle. The nearfield distributions corresponding to the normalized magnetic field intensity of first up-modulated frequency harmonic are also plotted in Fig. \ref{fig:Fig14}(b) for different incident angles. An almost perfect focusing is achieved in the off-axis directions set by the incident angle at the desired focal distance verifying the wide-angle performance of the focusing lens. The focusing efficiency is maximal for normal incidence and decreases by increasing the incident angle due to the decrement in frequency conversion efficiency, consistent with the results in Fig. \ref{fig:Fig4}. 

Further information on the power conversion to the frequency harmonics by the finite time-modulated metasurfaces vs incident angle can be found in the section 7 of Supplemental Material.

\section{Nonreciprocal Response and Isolation}
In this section, we look at one of the highly applauded features of space-time gradient metasurfaces which is the magnetless nonreciprocal response \cite{shaltout2015time,sounas2017non,lira2012electrically,hadad2015space,hadad2016breaking,correas2016nonreciprocal,taravati2017nonreciprocal,taravati2017mixer,chamanara2017optical,taravati2017nonreciprocal2,shi2017optical} and explore the utility of modulation-induced phase shift in obtaining a strong nonreciprocal response and isolation via spatial decomposition of frequency harmonics. The origin of nonreciprocity in space-time gradient metasurfaces is the difference in the imparted momentum by the metasurface to the forward and backward propagating waves. This comes to the picture when the light is carrying a momentum along the direction of spatiotemporal modulation which is the case in oblique incidence \cite{shaltout2015time,hadad2015space,shi2017optical}. In fact, the nonreciprocal response of the time-modulated metasurface under oblique incidence can be understood through Fig. \ref{fig:Fig12}, where by flipping the sign of incident angle, the received fields at the opposite observation angles will be of opposite frequency harmonic order.

\begin{figure}[htbp]
\centering
{\includegraphics[width=11.5cm]{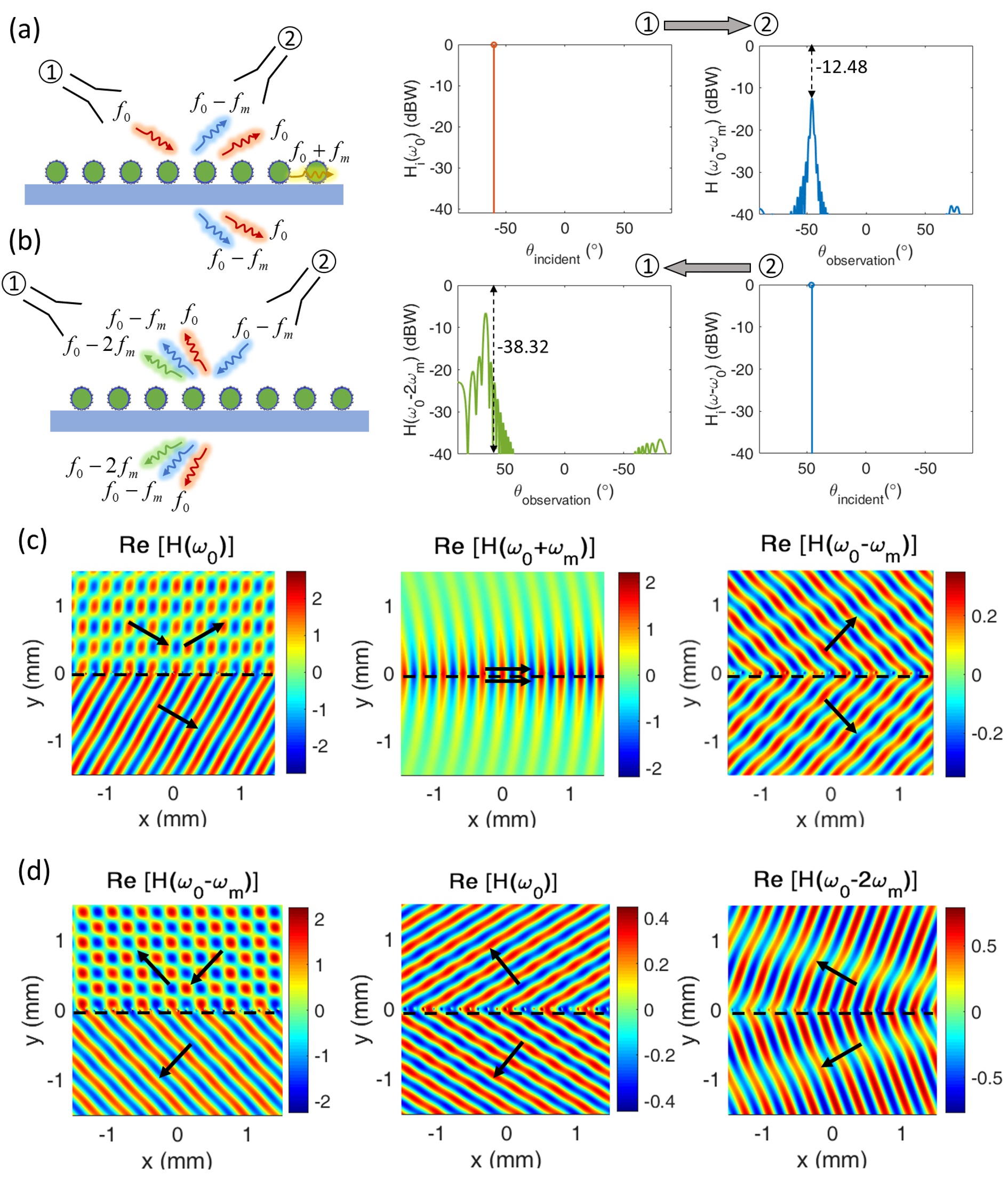}}
\caption{The nonreciprocal response of the time-modulated metasurface. (a) depicts the excitation of time-modulated metasurface from port 1 and reception at port 2 with the transmitted and received magnetic field intensities plotted calculated as functions of angle. (c) demonstrates the corresponding magnetic field wavefronts of the generated frequency harmonics to this link. (b) shows the excitation of time-modulated metasurface from port 2 under time-reversal and reception at port 1 where  the transmitted and received magnetic field intensities are calculated and plotted as functions of angle. (d) represents the corresponding magnetic field wavefronts of the generated frequency harmonics to this link.}
\label{fig:Fig15}
\end{figure}

In order to analyze the nonreciprocal response of the time-modulated metasurface in more depth, we choose the metasurface studied in Fig. \ref{fig:Fig12} and \ref{fig:Fig13} having a linear modulation delay profile with a progressive delay of $\Delta\alpha=k_0\Lambda\sin(10^{\circ})$. Since the degree of non-reciprocity and isolation depends on $\frac{f_m}{f_0}$, here we consider a larger modulation frequency of $f_m=30$ GHz which has been experimentally demonstrated for graphene \cite{phare2015graphene}. In particular, we are interested in large oblique incident angles for which the spatial symmetry between up- and down-modulated frequency harmonics is largely broken. First, we consider the metasurface excited by an obliquely incident TE-polarized plane wave at an angle of $\theta_{\text{incident}}=-60^{\circ}$ and wavelength of $\lambda_0=300$ $\mu$m from the port 1 at the left-hand side as depicted in Fig. \ref{fig:Fig15}(a). The magnetic field wavefronts of the fundamental and first-order frequency harmonics corresponding to this case are shown in Fig. \ref{fig:Fig15}(c). As it is observed in the results and can be verified from a simple calculation based on Eq. (22), the fundamental frequency harmonic is scattered toward $\theta_0=-60^{\circ}$, the first up-modulated frequency harmonic is guided along the metasurface with minimal radiation into free-space and the first down-modulated frequency harmonic is bent toward $\theta_{-1}\approx-45.5^{\circ}$ in the account of modulation-induced phase shift which is received at port 2. The transmitted magnetic field intensity from port 1 and the received magnetic field intensity at port 2 corresponding to this link are plotted in Fig. 15(a) in terms of dBW (decibels relative to one watt) as functions of angle. Next, we consider the metasurface under time-reversal i.e. the metasurface is excited by the frequency of $f_0-f_m$ incoming at an angle of $\theta_{\text{incident}}=+45.5^{\circ}$ from port 2 at the right-hand side, as depicted in Fig. \ref{fig:Fig15}(b). The magnetic field wavefronts of generated frequency harmonics up to the first-order corresponding to this case are plotted in Fig. \ref{fig:Fig15}(d). In this case, the fundamental harmonic of $f_0-f_m$ is scattered through ordinary refraction toward $\theta_{-1}=45.5^{\circ}$, while first up- and down-modulated frequency harmonics are bent toward $\theta_{0}\approx31.5^{\circ}$ and $\theta_{-2}\approx+66.25^{\circ}$, respectively due to the modulation-induced phase shift. Since $\theta_{0}$ and $\theta_{-1}$ are well-separated from $\theta_{\text{incident}}$, their transmission to port 1 is negligibly small. Therefore, under time-reversal, the received wave at port 1 will have a frequency of $f_0-2f_m$ which is different than the reciprocal frequency of transmitted wave ($f_0$) from port 1. This is a clear demonstration of the nonreciprocal response of the time-modulated metasurface which isolates the transmission and reception modes in both spatial and temporal domains by establishing nonreciprocal links. The transmitted magnetic field from port 2 and received at port 1 are plotted in Fig. 15(b) as functions of angle in terms of dBW. The separation between $\theta_{0}$ and $\theta_{-2}$ is proportional to the ratio of $f_m/f_0$ and increases by increasing the modulation frequency \cite{shaltout2015time}. Here, we have used a modulation frequency of $f_m=30$ GHz which has been show to be accessible experimentally and leads to an angular separation of $|\theta_{0}|-|\theta_{-2}|\approx6.25^{\circ}$. Due to the high directivity of the steered beams by the metasurface, a great isolation is achieved between ports 1 and 2 such that $S_{12}=-12.48$ dBW and $S_{21}=-38.32$ dBW as shown in Fig. 15 (a) and (b), which translates to an isolation level of 25.84 dBW.
\\

\section{Conclusion}

In conclusion, we introduced a design rule for designing time-modulated metasurfaces by changing the phase delay in modulation using phase shifters which allows for obtaining a dispersionless phase shift covering $2\pi$ span in the forward and backward scattering of generated frequency harmonics, regardless of incident angle and polarization. This principle can be used for designing efficient electrically tunable time-modulated metsurfaces to engineer the wavefronts of generated frequency harmonics. The applicability of the design principle is illustrated for steering and focusing the generated frequency harmonics in a time-modulated metasurface based on graphene-wrapped silicon microwires which is optically excited in the THz regime and electrically modulated in RF regime. Furthermore, the spectral and angular performance of the time-modulated metasurface in wavefront engineering is comprehensively studied to verify the constant phase response and the preserved functionality at generated frequency harmonics under incidence of different wavelengths and angles. The nonreciprocal response of time-modulated metasurface in wavefront engineering and utility of modulation-induced phase shift in creating isolators are also demonstrated. The proposed design paradigm is a clear departure from the previous tunable metasurface designs as it provides a full control over the phase modulation with a constant amplitude while offering a constant phase response versus incident wavelength and angle. As such, it holds a great promise for overcoming several major limitations facing active metasurfaces. 

It should be remarked that the proposed design rule and the physical discussions presented in this paper are applicable to any time-modulated metasurface with arbitrary-shaped unit cells incorporating different types of electro-optical materials such as graphene, transparent conducting oxides, doped semiconductors and transition metal nitrides. This allows for achieving dynamic manipulation of harmonics in different frequency regimes from visible to mid-infrared. Furthermore, there is a great potential for further improvement of frequency conversion efficiency and engineering the directionality by exploration of alternative geometries and realization of Huygens time-modulated metasurfaces.

\section*{Acknowledgements}
This work is supported by the US Air Force Office of Scientific Research (AFOSR), FA9550-14-1-0349 and FA9550-18-1-0354, in parts.
\bibliography{sample}

\end{document}